\documentclass[review]{elsarticle}
\usepackage{xfrac}
\usepackage{lineno,hyperref}
\usepackage{graphicx}
\usepackage{subcaption}
\usepackage{caption}
\usepackage{indentfirst}
\usepackage{siunitx}
\usepackage{amsmath}
\usepackage{float}
\usepackage{fullpage}

\makeatletter 
\def\ps@pprintTitle{%
 \let\@oddhead\@empty
 \let\@evenhead\@empty
 \def\@oddfoot{}%
 \let\@evenfoot\@oddfoot}
\makeatother

\usepackage{subcaption}			
\usepackage{geometry} 			
\usepackage{hyperref}			
\usepackage{eurosym}			
\usepackage{comment}			
\usepackage{amsmath}
\usepackage{amssymb}
\usepackage{amsfonts}
\usepackage{amsthm}
\usepackage{ulem}				
\usepackage[dvipsnames]{xcolor}	
\usepackage{todonotes}			
\usepackage{algorithm}			
\usepackage{algpseudocode}		
\usepackage{caption}
\usepackage{subcaption}			
\usepackage{multirow}			

\DeclareMathOperator*{\argmax}{arg\,max}	
\DeclareMathOperator*{\argmin}{arg\,min}

\title{Adaptive design of experiments methodology for noise resistance with unreplicated experiments}

\author[1]{Lucas Caparini}
\author[1]{Gwynn J. Elfring}
\author[1]{Mauricio Ponga\corref{correspondingauthor}}
\cortext[correspondingauthor]{Corresponding author}
\ead{mponga@mech.ubc.ca}
\address[1]{Department of Mechanical Engineering, 2054-6250 Applied Science Lane, Vancouver BC, Canada, V6T 1Z4}
\date{}

\begin{document}


\begin{abstract}
A new gradient-based adaptive sampling method is proposed for design of experiments applications which balances space filling, local refinement, and
error minimization objectives while reducing reliance on delicate tuning parameters.  High order local maximum entropy approximants are used for metamodelling,
which take advantage of boundary-corrected kernel density estimation to increase accuracy and robustness on highly clumped datasets, as well as
conferring
the resulting metamodel with some robustness against data noise in the common case of unreplicated experiments. Two-dimensional test cases are
analyzed against full factorial and latin hypercube designs and compare favourably. The proposed method is then applied in a unique manner to the
problem of adaptive spatial resolution in time-varying non-linear functions, opening up the possibility to adapt the method to solve partial differential equations. 

\end{abstract}

\maketitle


\section{Introduction}

Practical design problems often involve adjusting multiple parameters and understanding how each parameter influences quantities of interest. These
quantities essentially are functions of multiple variables and often display complicated behaviour. Generally, it is not realistic to model such
functions analytically, so numerous experiments, often numerical, are conducted to collect data and construct a simplified metamodel, from which
further design decisions can be made.

Design of experiments (DoE) techniques have been developed to create an optimal metamodel from affordable data points and are broadly categorized into
adaptive and non-adaptive techniques. Non-adaptive methods seek to increase the accuracy of the metamodel by placing samples in a statistically
optimal manner throughout the parameter space \cite{Pronzato2013} without using any additional information gained from previous sample points
\cite{Jin2002}. Once all desired points in the design space have been selected the data may be fit with any appropriate metamodel. Common examples
include full factorial designs (FF), latin hypercube (LH) sampling \cite{McKay1979}, and orthogonal arrays \cite{Owen1992}. A large family of
non-adaptive methods treat data point selection as a large-scale optimization problem with an objective function chosen to maximize information
entropy \cite{Shewry1987,Ko1995,Merkle1998}. The maximum-entropy (max-ent) approach has gained popularity because it minimally biases the metamodel,
and provides greater resistance to measurement error, but suffers computationally from being combinatorial in nature. Thus, the many gradual advances
in this approach have seen the non-deterministic polynomial-time-hard combinatorial optimization problem, which is generally not solvable in
polynomial time, steadily become tractable for larger problems \cite{Guest2009,Pronzato2013,Wang2021}. Other modern advances include features such as
Bayesian frameworks for increased performance on highly nonlinear functions \cite{Hamada2001}, handling of missing feature data \cite{Velicheti2021},
and the incorporation of sparsity-promoting heuristics, such as found in compressed sensing \cite{Diaz2018,Yu2018,Ibanez2019}. Nonetheless, the vast
majority of non-adaptive methods essentially result in a space-filling distribution of data points through the design space.

Previous works have elucidated some potential benefits of non-space-filling designs from a statistical perspective \cite{Pronzato2012}, validating the
intuition that some areas may need more refinement than others. Adaptive DoE methods utilize information from previous data points to recommend future
test points, theoretically allowing targeted refinement in areas where the metamodel poorly approximates the underlying function. Early efforts were
not met with consistent success \cite{Jin2002}, but that situation has been steadily improving. While some effort has been made to reconcile the
maximum-entropy family with fully adaptive design \cite{Youssef2019}, many recent efforts use common engineering heuristics instead of a rigid
statistical approach, and have shown impressive performance compared to space-filling methods on test problems \cite{Mackman2010, Li2010,
Theunissen2018}. Recent developments have further extended and improved these techniques by incorporating data-dependent smoothing parameters, which
improved results over noisy experimental data \cite{Theunissen2018}. Those adaptive methods which remain dominantly statistical will generally use
some approximation of metamodel error in the adaptivity scheme \cite{Azzimonti2021}. Recent works have particularly affirmed the usefulness of
leave-one-out (LOO) cross-validation error in adaptive DoE \cite{Li2018,Kyprioti2020}. At this point adaptive approaches have progressed to
demonstrate success on both analytical test problems and practical applications, such as chemical engineering \cite{Kaneko2021} and aerodynamics
\cite{DaRonch2017}. 

Unlike non-adaptive methods, in adaptive DoE methods the metamodel is an important component of the technique itself. Nearly all modern DoE techniques
must create metamodels over unstructured multidimensional data. The number of unstructured multidimensional interpolation techniques available is
relatively small despite active research in the field for many years \cite{Buhmann2003,Wendland2005}. Nearest-neighbor interpolation and inverse
distance weighting are easy to implement but have obvious and well-known drawbacks. Mesh generation difficulties limit popular engineering techniques
such as finite element shape functions even in three dimensions \cite{Board2019}, let alone arbitrary dimensions. Moving least squares and Radial
Basis Functions (RBF) require solving a costly system of equations and are relatively parameter-sensitive \cite{Liu2015}. Gaussian process regression,
also known as Kriging, is extremely popular among statistically-based adaptive DoE methods \cite{Picheny2010,Kaneko2021,Azzimonti2021}, but has
similar parameter sensitivity. The local maximum-entropy (LME) functions \cite{Arroyo2006b} are a relatively new class of approximants designed
specifically for solving PDEs with an unstructured multidimensional discretization \cite{Li2009a, Habbal2009}, and using only a single easily tuned
parameter. While they have seen much attention in the realm of meshfree PDE simulation \cite{Foca2015, Jiang2019, Wang2020}, little notice has been
taken outside of that field. Extensions of the basic LME formulation have resulted in a family of related approximants \cite{Cyron2009,Gonzalez2010}
of which the Higher-Order LME Scheme (HOLMES) \cite{Bompadre2012} is used in this work. 

The accuracy of the metamodel, and thus any adaptive DoE technique, depends on the reliability of the data provided to it. Multiple data points at the
same position in the design space can help alleviate this concern by providing some statistical bounds on the data values which can generate
data-dependent smoothing parameters for the metamodel \cite{Theunissen2018}. Unfortunately, even as sampling techniques and efficiency increase,
practical constraints frequently require the use of unreplicated data points from which no statistical features may be inferred \cite{Arat2016}. The
experimentalist is then forced to retreat to the delicate task of manually tuning metamodel smoothing parameters or forego them entirely and use an
exact interpolant.

This paper aims to take advantage of the natural error resistance of HOLMES approximants to create an adaptive DoE method that provides error
resistance without depending on multiple sensitive smoothing parameters. The manuscript is organized as follows: The adaptive DoE algorithm is
described in \autoref{sec:AdaptiveDoE}. Construction and use of HOLMES approximants in the metamodel is addressed in \autoref{sec:Metamodeling}, and a
simple kernel density estimation algorithm is proposed for automatically determining the local node spacing. The influence of the metamodel kernel
parameter on the novel DoE method is evaluated in \autoref{sec:Examples}, and the performance of the completed DoE algorithm is evaluated on a variety
of analytical solutions. Finally, \autoref{subsec:ADoEforTimeFuncs} sees the DoE method applied beyond the realm of experimental design, and used to
dynamically increase resolution of time-evolving functions.


\section{Adaptive Design of Experiments Scheme}
\label{sec:AdaptiveDoE}
Most DoE algorithms maximize or minimize some objective function when selecting new data points. Adaptive methods differ from non-adaptive by
utilizing the metamodel within the objective function, allowing the model to become iterative. This section details a specific choice of objective
function (\autoref{subsec:ObjFunc}) and associated iterative DoE algorithm (\autoref{subsec:PointSelect}). Details about the metamodel are left to
\autoref{sec:Metamodeling}.

	\subsection{Objective Function}
	\label{subsec:ObjFunc}
Assume a design space $\Omega \subseteq \mathbb{R}^d$, and an unknown function, $u: \Omega \rightarrow \mathbb{R}$, which may be evaluated through
either computer or physical experiments. Furthermore, assume $u(x)$ has already been evaluated at some points $x_a \in \Omega$ called nodes, or data
points. These values have been used to create a metamodel, $u_I(x;x_a) : \Omega \rightarrow \mathbb{R}$ which approximates $u$.

A scalar objective function, $S(x) : \Omega \rightarrow \mathbb{R}$, is desired which can be maximized to find the next data point for experimental
evaluation via
	\begin{gather} \label{eqn:NewPoint}
	x_{a+1} = \argmax_{x \in \Omega} S(x).
	\end{gather}

Several heuristics and practical constraints are here used to determine the form of $S(x)$. Assuming other factors are equal, one such heuristic
stipulates that proposed points should be placed in a space-filling manner throughout the domain. The associated constraint prevents duplicate
selection of data points. Similarly, areas of low metamodel accuracy -- where $u_I$ approximates $u$ poorly -- should garner more attention, while
those with perfect accuracy should not be suggested. A final heuristic follows the common intuition that areas of a domain where $u$ is highly
nonlinear require more data points to be well captured by $u_I$ and deserve more attention. Therefore, we propose the following heuristic objective
function in a simple separable form,
	\begin{gather} \label{eqn:ObjFunc}
	S(x) = Q_{L} Q_{S} Q_{E}.
	\end{gather}

\begin{figure}[h!]
\centering
\includegraphics[width=0.8\textwidth]{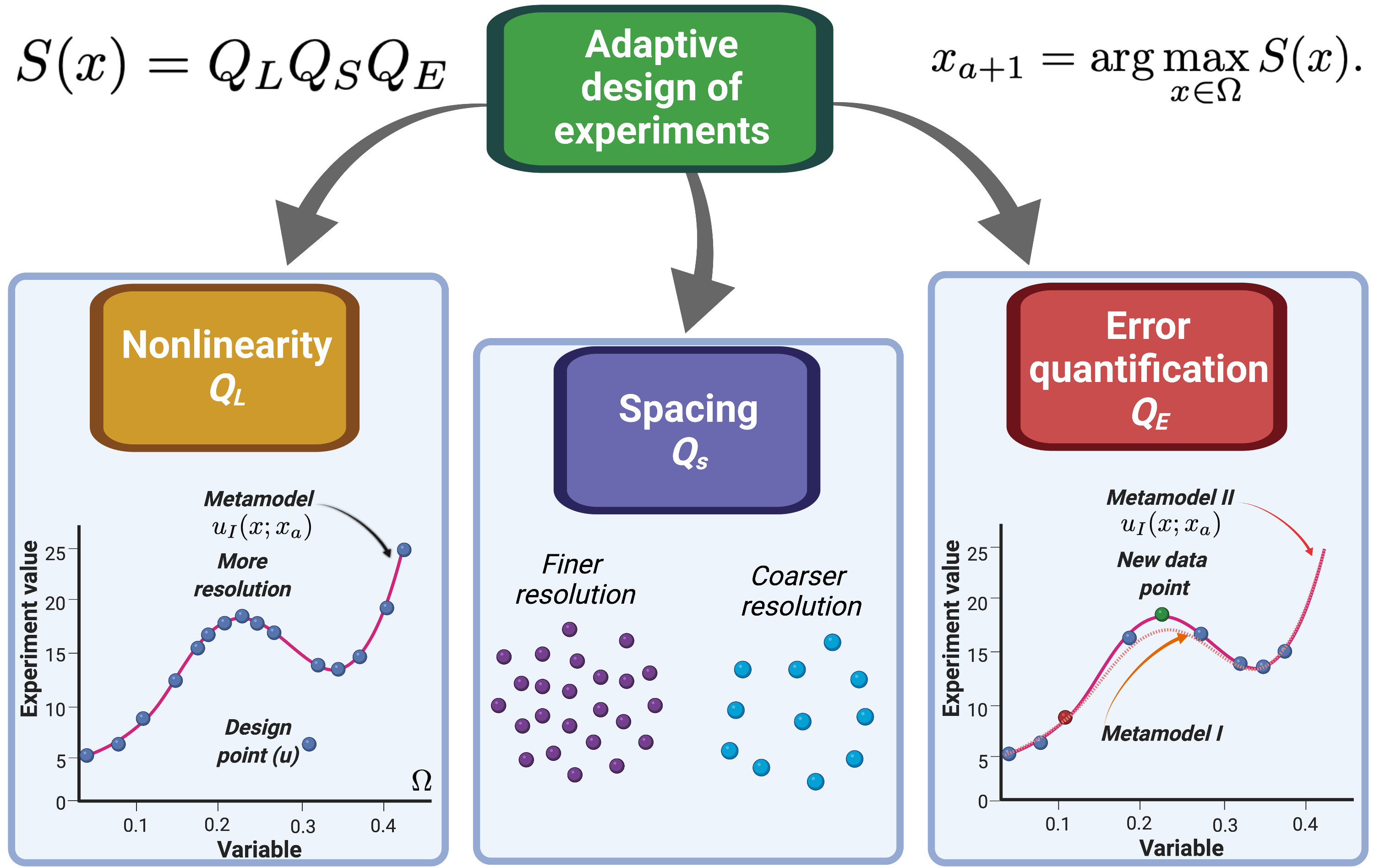}
\caption{Schematic representation of the adaptive design of experiment objective function proposed in this work. Three fundamental parts, namely, nonlinearity, spacing and error are key descriptors used in the heuristic approach. }
\label{Fig:DoESchematic}
\end{figure}

\autoref{Fig:DoESchematic} shows schematically the three different heuristics taken in our work. The objective function in \eqref{eqn:ObjFunc} is
composed of three multiplicative factors: $Q_L$ representing the contribution of nonlinearities, $Q_S$ the point spacing heuristic, and $Q_E$ the
metamodel error. Each factor is normalized onto the range $[\epsilon,1]$, where $\epsilon = 10^{-4}$ to prevent difficulties when one of the factors
becomes zero, or very close to zero, throughout the parameter space. In that case the resulting point placements would become dominated by numerical
error, rather than the remaining factors. In this case $S(x) : \Omega \rightarrow [\epsilon^3,1]$.

The product form of $S(x)$ is chosen to satisfy constraints associated with each heuristic by making $S(x)$ small when the constraint is violated. For
instance, if no metamodel error is observed in some area of the domain there is very little reason to place points in that region, so the the
objective function should assume a low value regardless of the other factors. Similar arguments can be made for the linearity and spacing components
of the objective function. If a component of the objective function needs to be weighted more than the others, its exponent can be altered, though the
effects of this were not heavily investigated in this work, and instead each component is treated as equally important. The objective function
proposed here is similar to other works \cite{Mackman2010, Li2010, Theunissen2018} in that a set of intuitive engineering heuristics are used to
select future data points. This form additionally avoids the combinatorially large optimization problem involved with compressed sensing
\cite{Brunton2019}, or non-adaptive max-ent DoE methods \cite{Bingham2015}. Each of the three factors is now expanded upon in detail.

		\subsubsection{Nonlinearity, $Q_L$}
		\label{subsubsec:Q_L}
The nonlinearity of the function, $u$, is emphasized in the objective function, $S$, based on the assumption that highly nonlinear areas will be
approximated more poorly than linear areas (see \autoref{Fig:DoESchematic}, left panel). The contribution is quantified by the magnitude of the
Laplacian of the metamodel,
	\begin{gather} \label{eqn:Laplacian}
	  s = |\nabla^2 u_I(x)|,
	\end{gather}
which is a convenient choice as an invariant of the Hessian matrix. This choice consciously prioritizes maxima and minima of functions, and would not
be ideal in the case of a perfect saddle point. If such a scenario was expected from the preliminary data, it would be simple to adjust this criterion
to use the sum of the magnitudes of the diagonal components of the Hessian, or some other simple metric. No significant changes were found in this
work with the examples selected, so the magnitude of the Laplacian is used through the remainder of this work.

To achieve consistent results with various functions, the Laplacian is normalized onto the range $[\epsilon,1]$ via
	\begin{gather} \label{eqn:Normalize}
		s_n(x) = \frac{s(x) - \min_{x\in\Omega} s(x)}{\max_{x\in\Omega} s(x) - \min_{x\in\Omega} s(x)}(1-\epsilon) + \epsilon.
	\end{gather}

Any quantity similarly normalized will be denoted $(\cdot)_n$ throughout the remainder of this work. Since \eqref{eqn:Normalize} only works when the
optimization is performed via an exhaustive search over $\Omega$, providing explicit values of $\min_{x\in\Omega}$ and $\max_{x\in\Omega}$. Other
optimization techniques for $S(x)$ must slightly alter this approach. Normalizations such as \eqref{eqn:Normalize} are performed using maximum and
minimum values from a previous iteration of the metamodel, and are observed to achieve similar results in the examples investigated here. Values of
$s_n(x)$ approaching computer precision are automatically rounded to zero to enforce consistent behavior of the algorithm.

The linearity factor, $Q_L$, requires an interpolation method to approximate second-order derivatives over scattered
data. $Q_L$ must be updated along with the metamodel and is selected to simply be the normalized Laplacian,
	\begin{gather} \label{eqn:Q_L}
	Q_L =  s_n.
	\end{gather}

	\subsubsection{Spacing, $Q_S$}
	\label{subsubsec:Q_S}
The spacing factor discourages clumping and prevents duplication of data points (see \autoref{Fig:DoESchematic}, central panel). The creation of a
suitable spacing factor was investigated thoroughly before \cite{Mackman2010} and the form 
	\begin{gather} \label{eqn:Q_S}
	Q_S = \left(1 - H(x) \right)^2
	\end{gather}
was decided upon after several tests.

Here, the spacing function $H(x)$ takes values of one at known data points and zero far away from data points. Mackman \textit{et al.}
\cite{Mackman2010} created a smooth $H(x)$ custom to the data set using Gaussian RBF interpolation with no appended polynomial constraints. The custom
spacing function is then a simple interpolation function using Gaussian functions centered at each node, as in 
	\begin{gather} \label{eqn:SpacingFunc}
		H(x) = \sum_{a = 1}^{N_a} \nu_a e^{-\xi\| x - x_a \|^2}.
	\end{gather}

Weights for each data point are obtained by solving the linear system 
	$$\sum_{j=1}^{N_a}\Phi_{ij}\nu_j = 1,$$
in which $N_a$ is the number of nodes, $\mathbf{\nu}$ is the unknown vector of weights, and the matrix is defined as $\Phi_{ij} = e^{-\xi\| x
- x_a \|^2}$ for $i,~j = 1, ~2, \cdots, N_a$.

Note that $Q_S$ must be updated after each point is proposed in order to prevent point duplication. 
	
The use of Gaussian RBFs requires the user to select a kernel parameter, $\xi$. This \textit{ad-hoc} selection has a significant influence on the
resulting points distribution and efficacy of the approach. The kernel parameter is selected based on a desired support radius, $R_{\text{supp}}$, and
numerical tolerance, $tol$,
	\begin{gather} \label{eqn:kernelfromfill}
		\xi = -\log\left(tol\right) / R_{\text{supp}}^2.
	\end{gather}
Previous works \cite{Mackman2010,Theunissen2018} achieved consistent results by selecting the desired support radius as a multiple of the fill
distance, $d_{fill}$, which is the largest nearest-neighbor distance:
	\begin{gather} \label{eqn:suppradfromfill}
		R_{\text{supp}} = R_0 \left( \frac{1}{2} d_{fill} \right).
	\end{gather}
Here $R_0$ is the scalar parameter and $d_{fill}$ is the fill distance given by
	\begin{gather}
		d_{\mathrm{fill}} = \sup_{x \in \Omega} \min_{x_a \in \mathcal{X}} \|x-x_a\|.
	\end{gather}
where $\mathcal{X}$ is the set of all nodes. The sensitivity of the algorithm to the parameter $R_0$ and suggestions for selection are detailed in
\autoref{subsec:KernelParam}.

	\subsubsection{Error Quantification, $Q_E$}
	\label{subsubsec:Q_E}

Leave-one-out cross-validation error (LOO) is well known within the statistical community for its superior error estimation abilities
\cite{Molinaro2005}. Its inclusion has significantly increased the performance of adaptive DoE methods in the past \cite{Kyprioti2020}. LOO quantifies
how much the addition of a data point to the model has increased the model's accuracy by comparing metamodel performance without the data to the known
value. To evaluate LOO at the known data point $x^\ast$, where $u(x^\ast)$ is known, a metamodel is constructed excluding this point, $u_I^\ast(x)$.
The LOO error is then given by comparison of $u_I^\ast(x)$ with $u(x)$ at $x^\ast$:
	\begin{gather} \label{eqn:LOO}
	\epsilon_\mathrm{LOO}(x^\ast) = \vert u(x^\ast) - u_I^\ast(x^\ast) \vert.
	\end{gather}

A point whose additional significantly influences the metamodel will garner a large LOO value. Additional sampling is likely needed in that vicinity,
since the metamodel is now known to have been capturing it poorly. This idea is schematically illustrated in \autoref{Fig:DoESchematic}, right panel.
The red point adds little improvement to the metamodel while the green one critically improves the metamodel. Therefore, the adaptive DoE technique
should be biased to further explore the area around the green point.

In our implementation, LOO errors at every data point are evaluated and subsequently normalized to the range $[\epsilon,1]$, with values approaching
computer precision receiving a value of zero automatically.

While LOO error can only be evaluated at known data points, $Q_E$ must be evaluated at any location in the design space, which is accomplished by
interpolation of LOO values. Here, the same interpolation method used for the metamodel is used for LOO error, with the implied assumption that
$\epsilon(x)_\mathrm{LOO,n} : \Omega \rightarrow [\epsilon,1]$ is a smooth function in space. Under this assumption, then we can evaluate $Q_E$ as
	\begin{gather} \label{eqn:Q_E}
	Q_E(x) = \epsilon_\mathrm{LOO,n}(x) = \sum_{i=1}^{N_a} w_{a}(x) \epsilon_\mathrm{LOO,n}(x_a),
	\end{gather}
where $w_a$ are the shape functions associated with the nodes $x_a$. Since $Q_E$ relies on explicit interpolation of known data points, it is updated
with the metamodel.

Evaluating LOO error requires reconstructing the metamodel at each node location for the evaluation of $u_I^\ast(x^\ast)$. If the metamodel is
challenging to construct or the dataset is large, this step can become costly. The metamodel described in \autoref{sec:Metamodeling} performs single
point evaluations quite quickly, so it is not significantly penalized by using LOO. The RBF metamodel used by others \cite{Mackman2010} is not a
reasonable choice as a metamodel in this scenario because of the high computational cost of metamodel construction which would be required with each
$u_I^\ast(x^\ast)$ evaluation.

Like with the linearity factor, $Q_L$, the LOO criterion, $Q_E$, cannot be reevaluated until new experimental data is collected. This allows multiple
points to be suggested at once by the algorithm without the computational cost of LOO evaluation. Furthermore, additional data points only require LOO
to be reevaluated in the area immediately surrounding them, rather than for the entire domain. This is described further in \ref{subsec:PointSelect}
below.

	\subsection{Point Selection Algorithm}
	\label{subsec:PointSelect}

A single data point may be found through equation \eqref{eqn:NewPoint}; however, most experimentalists will prefer receiving multiple data point
recommendations simultaneously to take advantage of batch processing. The proposed DoE algorithm can recommend multiple distinct data points to the
experimentalist, provided $Q_S$ is updated appropriately.

Suppose an experimentalist can gather $N_p$ data points per day, and would like to use the DoE algorithm to propose the next data points to gather
within the design space. There are $N_{iter}$ days allotted for collecting data, and a starting data set of $x_0$ data points and the corresponding
results of $u_0 = u(x_0)$. The resulting DoE and experimental workflow would look as in algorithm \ref{alg:DoE}. Each day the new data points would be
used to construct a new metamodel, from which $Q_L$ and $Q_E$ are derived. From there $N_p$ unique points are proposed by alternately evaluating
\eqref{eqn:NewPoint} and updating $Q_S$ to include the point already suggested. After all $N_p$ points have been recommended, the experimentalist can
add them to the dataset.

The described algorithm is necessary to recommend multiple points at once because $Q_S$ must be updated to prevent the same point being recommended
every time. Because $Q_S$ is not dependent on the experimental results, but only the position of previous points within the design space, it is
perfectly acceptable to update it without new experimental results. Similarly, neither $Q_L$ nor $Q_E$ can be adjusted until new experimental data is
available. There is no need to recompute them, and some computational power can be saved.

$Q_L$ and $Q_E$ cannot be updated when multiple points are proposed, but only whenever new experimental values are learned and the metamodel is
updated. This suggests a two-tiered update mechanism for the DoE algorithm. An inner loop will propose $N_p$ new data points by updating $Q_S$ at
every iteration. An outer loop takes the proposed points, evaluates them through experiements, updates the metamodel and associated $Q_L$ and $Q_E$
values, and finally passes everything back to inner loop to get more points. The outer loop operates for a specified amount of iterations,
$N_{\text{iter}}$, while would usually correlate to a predetermined number of data points or metamodel accuracy.

\begin{algorithm}[h!]
\caption{Adaptive DoE Procedure}
\label{alg:DoE}
	\begin{algorithmic}[1]
		\State Create initial data points, $x_0$.
		\State Evaluate $u_0 = u(x_0)$.
		\For{$i = 1, \ldots, N_\mathrm{iter}$}
			\State Create interpolation scheme, $u_I(x) = \sum_a w_a(x) u_{0,a}$.
			\State Evaluate and normalize LOO error at data points, $\epsilon_\mathrm{LOO,n}$.
			\State Interpolate $Q_L$ and $Q_E$ onto test points in domain.
			\For{$j = 1 ... N_{p}$}
				\State Create spacing function, $H(x_0,x_\mathrm{new})$.
				\State Interpolate $Q_S$ onto test points in domain.
				\State Select new optimum, $x_\mathrm{new,j} = \argmax S(x)$, through any optimization method.
			\EndFor
			\State Evaluate $u_\mathrm{new} = u(x_\mathrm{new})$.
			\State Concatenate Data, $x_0 \gets [x_0;x_\mathrm{new}], ~u_0 \gets [u_0;u_\mathrm{new}]$.
		\EndFor
	\end{algorithmic}
\end{algorithm}

The solution of equation \eqref{eqn:NewPoint} required for each iteration of the inner loop could be done in any convenient method. Every example in this
work has performed an exhaustive search over a finely discretized domain. More advanced methods will require small modifications to the normalizations
used in $Q_L$ and $Q_E$, as discussed in \ref{subsubsec:Q_L}.


\section{HOLMES Metamodel}
\label{sec:Metamodeling}

Evaluating the nonlinearity parameter \eqref{eqn:Q_L} requires a metamodel capable of approximating the Hessian of $u(x)$. The HOLMES approximants
developed by Bompadre \textit{et al.} \cite{Bompadre2012} are used here rather than the more common RBF, or polyharmonic spline (PHS) interpolation
methods. HOLMES is based on the LME shape functions \cite{Arroyo2006b} and ultimately the max-ent functions previously introduced by Sukumar
\cite{Sukumar2004}. HOLMES is known to function in any dimension without alteration to its original formulation, and does not require structured input
data. It also does not require the solution of a large system of equations during its evaluation. This section outlines the construction of HOLMES
and modifications necessary for use in the adaptive DoE method.

	\subsection{Basic Implementation}
	\label{subsec:BasicHOLMES}

The HOLMES consists in a set of shape, or basis, functions, $w_a(x)$, centered around a corresponding set of nodes, or data points, $x_a$, situated in
the design space $\Omega \subseteq \mathbb{R}^d$. If the values of the unknown function, $u(x)$, are known at the nodes, HOLMES may be used to
construct a metamodel in the same manner as any interpolation scheme,
	\begin{gather} \label{eqn:Interpolation}
		u_I(x) = \sum_{x_a \in \mathcal{N}_a(x)} w_a(x) u(x_a).
	\end{gather}
Here, $\mathcal{N}_a(x)$ denotes the set of nodes within a neighborhood of the query point $x$.

The HOLMES shape functions are formulated by solving a nonlinear multi-objective optimization problem which seeks to maximize both information entropy
and minimize shape function width. Additional constraints are added in the form of Lagrange multipliers to ensure $n^{th}$ order polynomial
consistency. Details of the derivation may be found in the original work \cite{Bompadre2012}.

The resulting functions are the sum of negative and positive exponential components,
	\begin{align}
		w_a(x) &= w_a^+(x) - w_a^-(x) \label{eqn:HOLMESdefa} \\
		w_a^\pm(x) &= \exp \left[-1 -h^{-p} \gamma \|x-x_a\|_p^p \mp \left( \sum_{\alpha \in \mathcal{A}_{d,n}} h^{-|\alpha|} \lambda_{\alpha} (x-x_a)^\alpha \right) \right]. \label{eqn:HOLMESdefb}
	\end{align}
The exponents are each composed of a simple decay term and an additional term which utilizes a Lagrange multiplier, $\lambda$, to explicitly enforce
polynomial consistency of order $n$ on the interpolant. The decay term, $-h^{-p} \gamma \|x-x_a\|_p^p$ uses a non-dimensional parameter, $\gamma$, the
$p-$norm of distance, $\|\cdot\|_p$, and a measure of the nodal spacing, $h$, to determine the width of the shape function.

The third term in the argument of \eqref{eqn:HOLMESdefb} is centered on the Lagrange multiplier, $\lambda_{\alpha}$. This term uses the same $h$ value
as in the decay term to scale the Lagrange multiplier to roughly $\mathcal{O}(1)$ for faster convergence. Multi-index notation is used to concisely
write out the polynomial component terms. In this notation, $\alpha \in \mathbb{N}^d$ is a multi-index used to represent each monomial component of an
$n^{th}$ order polynomial in $\mathbb{R}^d$. For instance, in three dimensions if $x = \begin{bmatrix}\mu_1&\mu_2&\mu_3\end{bmatrix}$ and $\alpha =
\begin{bmatrix}1&2&1\end{bmatrix}$, then the notation $x^{\alpha} = \mu_1\mu_2^2\mu_3$. $\mathcal{A}_{d,n}$ is then the set of all valid multi-indexes
combinations of dimension $d$ and order less than or equal to $n$. The Lagrange multiplier, $\lambda$, is a vector whose length corresponds to the
number of monomial components in a polynomial of the specified order and dimension, which can be easily found with the binomial coefficient function,
{\small $D_{d,n} = \begin{pmatrix} d+n\\ n \end{pmatrix}$}.

\subsubsection{Regularized Newton Iterations} \label{subsubsec:RegNewton}

The Lagrange multipliers must be found implicitly as the solution to a nonlinear convex minimization problem, i.e.,

		\begin{align} \label{eqn:OptLambda}
		\lambda^\ast(x) &= \argmin_{\lambda \in \mathbb{R}^{D_{d,n}}} Z(x,\lambda), \\
		Z(x,\lambda) &= \lambda_0 + \sum_{a = 1}^{N_a} \left[ w_a^+(x,\lambda) + w_a^-(x,\lambda) \right].
		\end{align}

The problem given by \eqref{eqn:OptLambda} can be solved through Newton-Raphson iterations with explicit formulas provided
by Bompadre \textit{et al.} \cite{Bompadre2012} and summarized in \eqref{eqn:OrigNewton1} through \eqref{eqn:OrigNewton2}. Here the vector
$\mathbf{r}$ is the gradient of $Z$ with respect to $\lambda$, and the matrix $\mathbf{J}$ is the Hessian with respect to $\lambda$. The multiindices
$\alpha,\beta \in \mathcal{A}_{d,n}$ correspond to specific components of the Lagrange multiplier vector, which is indexed by them.
		\begin{align}
		\label{eqn:OrigNewton1}
		r_0(x,\lambda) &= \frac{\partial Z(x,\lambda)}{\partial \lambda_0} = 1-\sum_{a=1}^{N_a} \left(w_a^+(x,\lambda) - w_a^-(x,\lambda)\right),\\
		r_{\alpha}(x,\lambda) &= \frac{\partial Z(x,\lambda)}{\partial \lambda_{\alpha}} = -h^{-|\alpha|}\sum_{a=1}^{N_a} \left(w_a^+(x,\lambda) - w_a^-(x,\lambda)\right)\left(x-x_a\right)^{\alpha},\\
		J_{\alpha\beta} &= \frac{\partial^2 Z(x,\lambda)}{\partial \lambda_{\alpha} \partial \lambda_{\beta}} =
		h^{-(|\alpha|+|\beta|)} \sum_{a=1}^{N_a}\left(w_a^+(x,\lambda) + w_a^-(x,\lambda)\right)\left(x-x_a\right)^{\alpha + \beta}.
		\label{eqn:OrigNewton2}
		\end{align}

In practice, the Hessian matrix, $\mathbf{J}$, is often poorly conditioned -- an effect which becomes worse as $n$ or $d$ increase. This issue
imposes a practical limit to the order and dimension of HOLMES evaluations.

For the purposes of the present work, a regularization proposed by Polyak \cite{Polyak2009} suitably resolves this issue. It is implemented via a
modification of the partition function, $Z(x,\lambda)$, into $\hat{Z}(x,\lambda,\Lambda)$, as seen in \eqref{eqn:ModNewton1} to
\eqref{eqn:ModNewton2}. This is a convenient modification because when the partition function and its derivatives are evaluated at $\Lambda = \lambda$
it returns the original partition function and gradient. The only values effected are the diagonal elements of the Hessian, to which the norm of the
gradient is added.

		\begin{align}
			\label{eqn:ModNewton1}
			\hat{Z}(x,\lambda, \Lambda) &= Z(x,\Lambda) + \frac{1}{2}\|\nabla Z(x,\lambda)\| \| \Lambda - \lambda \|^2, \\
			\hat{Z}(x,\lambda,\Lambda)|_{\Lambda=\lambda} &= Z(x,\lambda), \\
			\frac{\partial \hat{Z}(x,\lambda,\Lambda)}{\partial \Lambda} |_{\Lambda=\lambda} &= \mathbf{r}(x,\lambda), \\
			\frac{\partial^2 \hat{Z}(x,\lambda,\Lambda)}{\partial \Lambda_\alpha \partial \Lambda_\beta} |_{\Lambda=\lambda} &=
			\mathbf{J}(x,\lambda) + \| \mathbf{r}(x,\lambda) \|\mathbf{I} .
		\label{eqn:ModNewton2}
		\end{align}

The regularization above proved robust in all examples, but convergence issues appear again as $d$ and $n$ increase. Combining Newton's method with an
alternate optimization algorithm, such as Nelder-Mead or preconditioned conjugate gradient has been used to overcome similar difficulties with the
related LME shape functions \cite{Navas2018, Fan2018} but were not pursued here. Instead, effort was placed in adjusting the kernel width to better
suit the local node distribution and increase both robustness and accuracy. This is detailed in the next section.

	\subsection{Adjusting the Kernel Width}
	\label{subsec:AdjustingKernelWidth}
Any kernel-based interpolation scheme may face practical computational issues on real data sets. One such issue occurs when the data distribution is
extremely uneven. Clumps of high density data occur in some places, while others are left relatively sparse. In theory, such clumped datasets would
not cause an issue for HOLMES, but in reality it may lead to difficulty solving the convex optimization problem required for evaluating the Lagrange
multipliers. Since any adaptive DoE approach will result in unevenly spaced datasets and potential clumping, it is worth investigating the effects an
adjustable kernel parameter may have on HOLMES in terms of robustness and accuracy.

		\subsubsection{Nodal Spacing Parameter}
		\label{subsubsec:NodalSpacingParameter}
Before any method of adjusting the kernel parameter can be introduced or evaluated, HOLMES must first be implemented in a manner that accepts variable
kernel parameters. A similar operation has been performed by for the related LME shape functions \cite{Arroyo2009}. Though this is a small
step, it is extremely important when considering situations where very large differences in spacing may be desirable. A
constant kernel width is simply not feasible in such cases. Section \ref{sec:Examples} below will explore applications which could potentially result in
such situations.

In equation \eqref{eqn:HOLMESdefb} the kernel parameter is expressed by the term $\gamma h^{-p}$, where $\gamma$ is non-dimensional and $h$ is a
measure of the nodal spacing. Na\"ively, $h$ could be adjusted dynamically as a function of the query point location, $h(x)$, but this introduces
additional terms into the HOLMES derivatives which are difficult to evaluate without an explicit formula for $h(x)$. Instead, the approach developed
by Ref. \cite{Arroyo2009} is taken. Instead of a single $h$ value being used for all nodes, each node, $x_a$, is associated with its own $h_a$ value.
These $h_a$ values must either be constant or determined in some other way which does not depend on the query point location in order to avoid
generating more complicated derivative formulations for HOLMES.

This approach can be used with HOLMES provided one crucial observation: \textit{Lagrange multipliers must only be a function of $x$}. The scaling term
$h^{-|\alpha|}\lambda_{\alpha}$ only exists to speed convergence of Newton's method and scale the elements of $\lambda$ approximately onto the range
$[0,1]$. In this role, $h$ has no actual influence on the final shape function. Allowing $h$ to be a nodal parameter does not make sense inside this
term and will ruin HOLMES convergence. The related LME literature never encountered such issues because, with the exception presented in Ref.
\cite{Foca2015} where adjustable $h_a$ was not used, $\lambda$ was never normalized by $h$. The nodal spacing parameter used for this scaling must now
be distinct from the one used within the kernel parameter, and will be denoted $h_g$. It can be selected as the global average node spacing or as a
function of position with a little care, but can never be a nodal parameter.

To further differentiate the node-dependent $h$ value used in the kernel parameter, $\gamma h_a^{-p}$, from the $h$ value use to scale the Lagrange
multipliers, $h_g$, the substitution $\beta_a = \gamma h_a^{-p}$ is made within the HOLMES shape functions. $\beta_a$ represents the scaled,
dimensional kernel parameter of node $x_a$, just as in the previous LME literature \cite{Arroyo2009}. The new flavor of HOLMES is given by equation
\eqref{eqn:newHOLMES},
\begin{gather} \label{eqn:newHOLMES}
w_a^\pm(x) = \exp \left[-1 -\beta_a\|x-x_a\|_p^p \mp \left( \sum_{\alpha \in \mathcal{A}_{d,n}} h_g^{-|\alpha|} \lambda_{\alpha} (x-x_a)^\alpha \right) \right].
\end{gather}

Unlike the method used in Ref.~\cite{Arroyo2009}, our modification does not substantially change the original derivative formulations, making the
substitution quite convenient. Determination of the local average spacing is the subject of the next section.

		\subsubsection{Kernel Density Estimation}
		\label{subsubsec:KDE}
The spacing parameter $h_a$ can now be chosen to reflect the local nodal spacing around a given point, rather than the global average. The approach we
take is to approximate the local node
density, $\rho(x)$ at a node, and deduce the local average spacing through
	 \begin{gather} \label{eqn:ha}
	   h(x) = \frac{1}{\rho(x)^{1/d}},
	 \end{gather}
since a density value is more intuitive than an average spacing value.

Kernel density estimation (KDE) has been used by the statistical community for many years to reconstruct an approximate probability density function
from binned data \cite{Rosenblatt1956}. KDE is a kernel convolution method with straightforward implementation and extensive literature
\cite{Silverman1998}. To see how it is applied to non-binned point data in the form of nodal positions, consider convolving some density function,
$\rho(x)$ with some kernel, $K(x)$, in order to obtain a smoothed approximation of the density function, $\hat{\rho}(x)$,
		\begin{gather} \label{eqn:convolution}
		\hat{\rho}(x) = \int_{-\infty}^{\infty} \rho(x) \cdot K(x-y) dy.
		\end{gather}
If the kernel, $K(x)$, is chosen appropriately, it is known that this approximation becomes exact as the kernel width decreases to zero
\cite{Silverman1998,Liu2010}. However, the function $\rho(x)$ is not known beforehand except for at selected nodal points. Approximate the number
density at each node by the global density over the domain, $\rho(x_a) \approx \rho_{avg} = \sfrac{N_a}{V_{\Omega}}$, and approximate the integral as
	\begin{gather} \label{eqn:convolution2}
	  \hat{\rho}(x) \approx \rho_{avg} \sum_{a=1}^{N_a} K(x-x_a) \Delta V_a,
	\end{gather}
where, like the density at the nodes, the volume associated with each node is taken as the global average, $\Delta V_a = \sfrac{V_{\Omega}}{N_a}$.
Since $\rho_{avg}\Delta V = 1$, the density estimate becomes a simple sum of kernel functions,
	\begin{gather} \label{eqn:DensityApprox}
	  \hat{\rho}(x) \approx \sum_{a=1}^{N_a} K(x-x_a).
	\end{gather}
Thus, an estimate of node density can be evaluated at a point by a simple summation of kernel functions centered around adjacent nodes.

The simplicity and extensive history of KDE makes it an obvious choice for node density estimation. In fact, some LME works used a \textit{local
average} method to find $h(x_a)$,
		\begin{gather} \label{eqn:old_ha}
		  h_a = \left(\frac{V_{k,a}}{k}\right)^{1/d},
		\end{gather}
which can be viewed as KDE with a rectangular window kernel whose size adjusts to encompass $k$ neighbours. Here $V_{k,a}$ is the volume of
hypersphere encompassing $k$ nodes around point $x_a$, and $d$ is the spatial dimension.

		  \begin{figure}[h]
		  \centering
		    \centering
		    \includegraphics[width=0.8\textwidth]{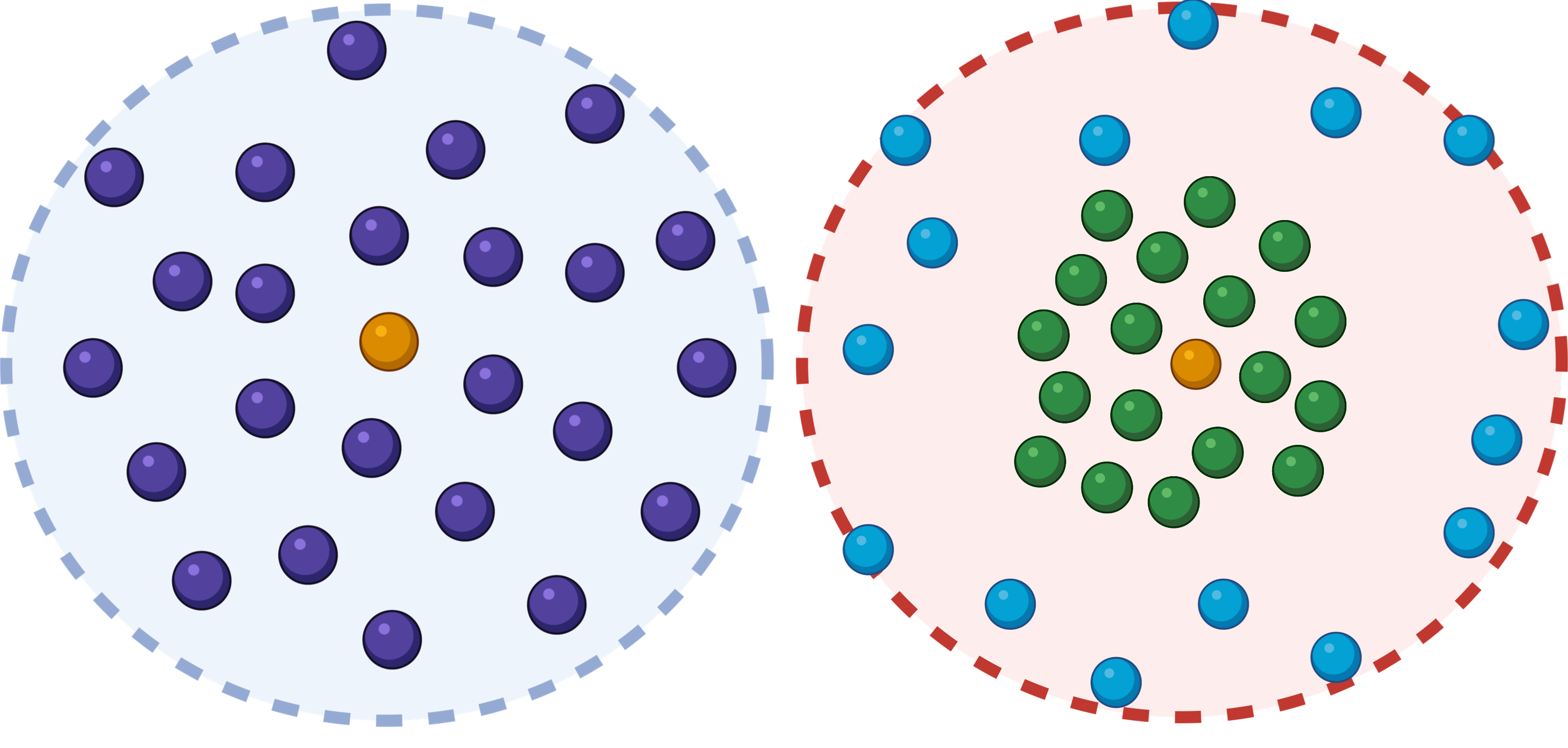}
		    \caption{Smoothly changing point density \textit{vs.} clustered data distribution. The points in the left figure are homogeneously distributed near the orange data point where as the green points shown in the right image are clustered around the orange ones. Moreover, the light blue points are placed further away from the orange point clearly indicating a change in the spatial resolution. }
		    \label{Fig:Point_Dist:smooth}
		  \end{figure}

A rectangular kernel will work well when the point density is nearly constant or changes slowly through the domain as shown in the left illustration
in \autoref{Fig:Point_Dist:smooth}, but will be poor when clustering exists, as in the right illustration in \autoref{Fig:Point_Dist:smooth}. Instead,
a more suitable choice of the kernel is to weight points in a smooth and monotonically decreasing manner from the node. Using a Gaussian function to
approximate nodal density seems like the most obvious choice,
	\begin{gather} \label{eqn:dens_approx_Gauss}
	  \hat{\rho}(x) \approx A \sum_{x_a \in \mathcal{N}_{a}(x)} \exp{ \left[-\frac{(\mathbf{x}_a-\mathbf{x})\Sigma (\mathbf{x}_a-\mathbf{x})}{2} \right] },
	\end{gather}
	\begin{gather} \label{eqn:Norm0}
	  A = \frac{1}{\sqrt{(2\pi)^{d}| \Sigma |}}.
	\end{gather}
Here $\Sigma$ is a matrix of decay parameters, which in this case is chosen to be diagonal such that $\Sigma_{ii} = \sigma_i^2$ for some vector
$\sigma \in \mathbb{R}^d$. The constant $A$ is chosen to normalize the integral. The Gaussian is a good choice because it offers well-studied
properties and takes few parameters, while also sporting much faster tail decay than simpler alternatives such as constant or linear kernels. Rapid
tail decay is especially important for effective KDE estimation in higher dimensions \cite{Silverman1998}. It is also smoother than other popular
options such as the Epanechnikov kernel \cite{Epanechnikov1969}. Finally, while more specialized kernels often require a rapidly increasing number of
parameters as the dimension increases, making them cumbersome for unsupervised tasks, the Gaussian kernel is trivial to apply to arbitrary dimensions.

\begin{figure}[h!]
	  \centering
	  \begin{subfigure}[b][][t]{0.415\textwidth}
	    \centering
	    \includegraphics[width=\textwidth]{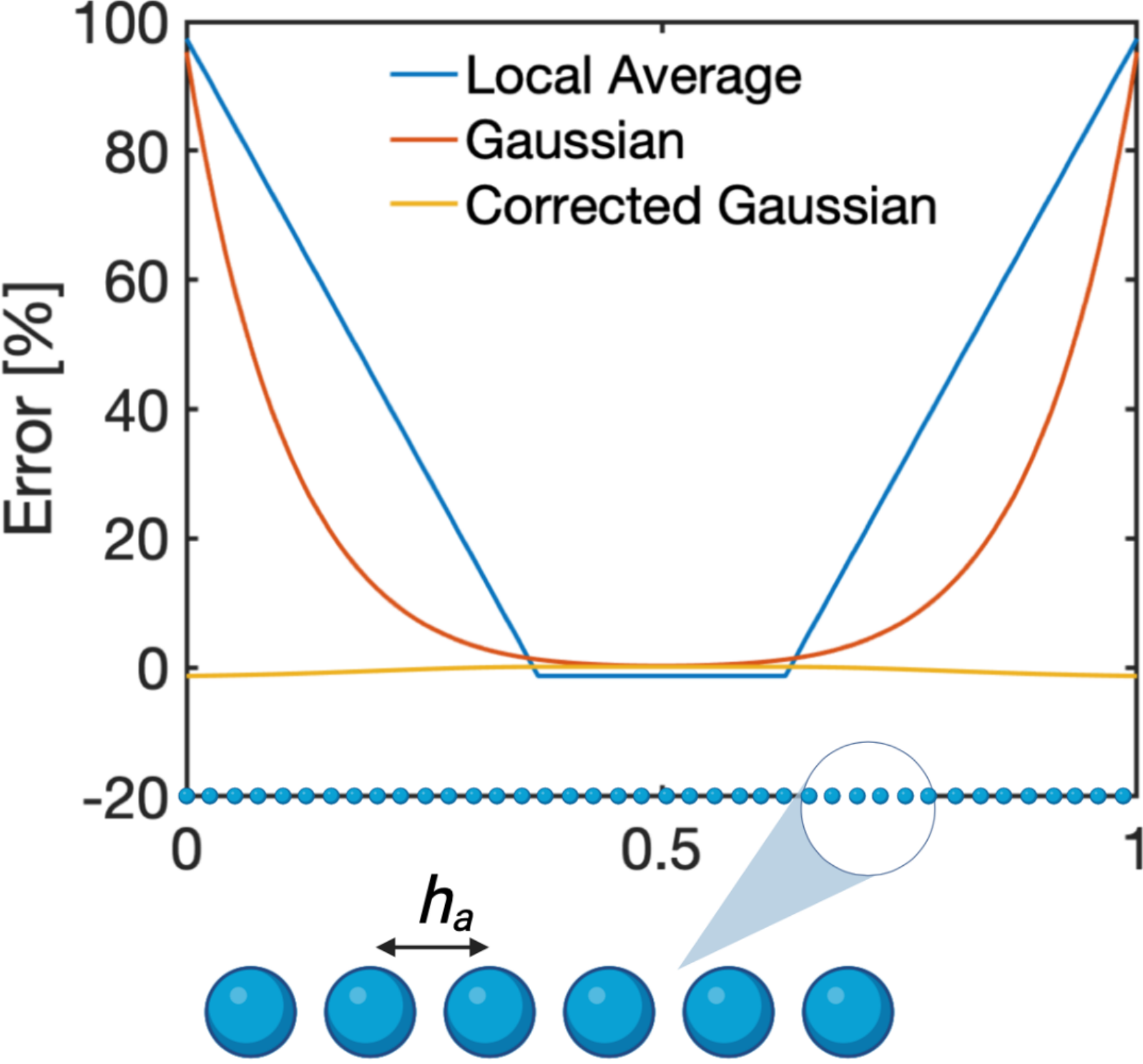}
	    \caption{Error in $h_a$ approximation.}
	    \label{Fig:ErfInfluence}
	  \end{subfigure}
	  \hfill
	  \begin{subfigure}[b][][t]{0.545\textwidth}
	    \centering
	    \includegraphics[width=\textwidth]{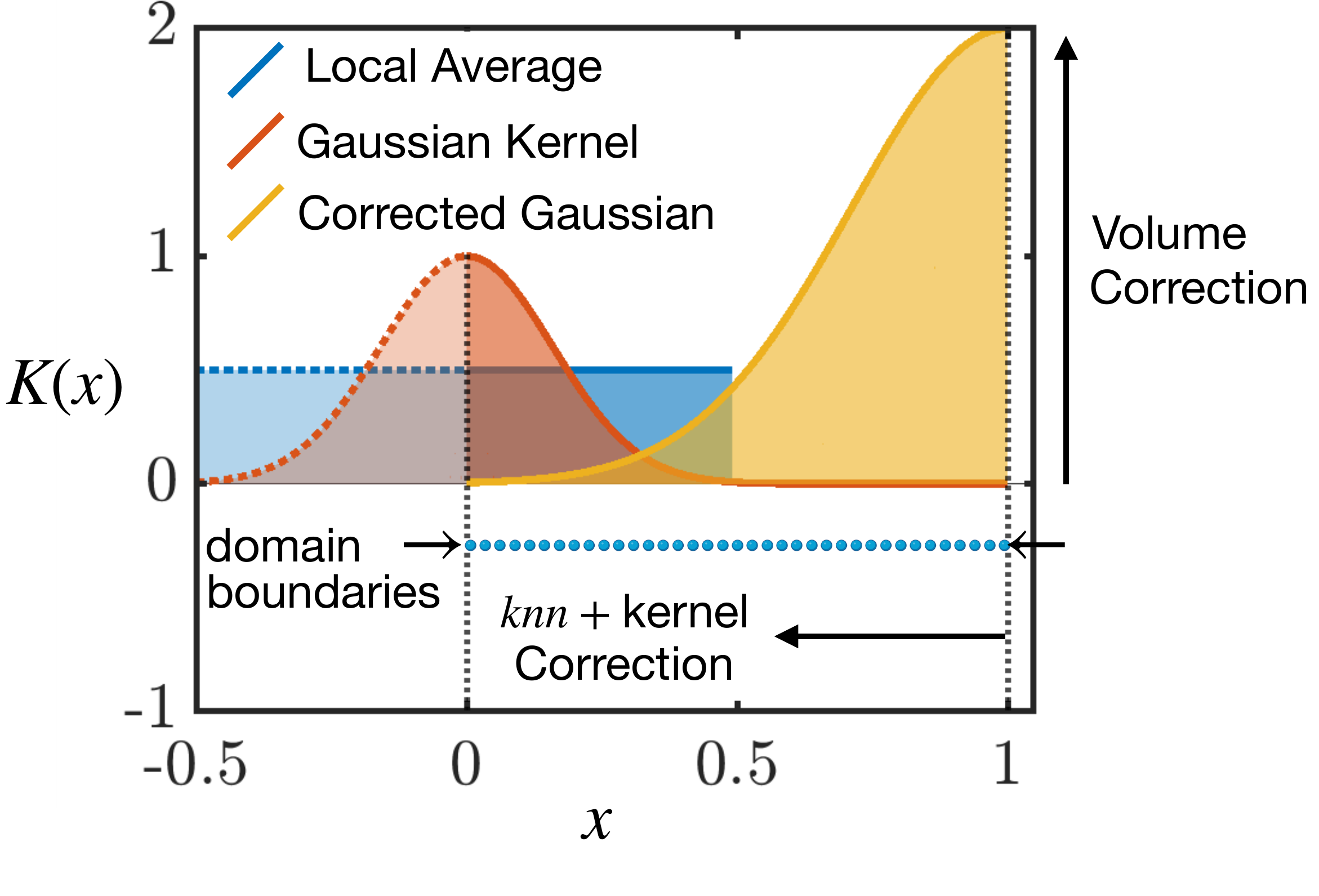}
	    \caption{Schematic representation of the different level of corrections used for estimating $h(x)$.}
	    \label{Fig:AdjustedComparison}
	  \end{subfigure}
	  \caption{Behaviour of several boundary-corrected kernel density estimation method on 1D line with constant nodal spacing $h_a = 0.01$ using $k=75$ nearest neighbours to estimate $h(x)$. (a) Error in $h_a$ \textit{vs.} $x$ for local average \eqref{eqn:ha}, Gaussian \eqref{eqn:dens_approx_Gauss} and three-step boundary correction procedure. (b) Schematic representation of the different approaches used for comparison. It is evident that the local average and Gaussian kernel try to retrieve information outside of the domain boundary were there is no data (the local average and Gaussian extend beyond the left boundary placed at $x=0$. The corrected Gaussian approach rescales the Gaussian to avoid retrieving information outside the boundaries and thus, the Gaussian is rescaled.}
	  \label{Fig:haPerformance1d}
	\end{figure}

The accuracy of KDE for this application is here assessed by comparison to a set of points placed at a known constant spacing. Let $\Omega = [0,1]$
and an even nodal spacing of $h_a=0.01$ be used to create a node-set within $\Omega$. \autoref{Fig:ErfInfluence} demonstrates the poor performance of
the local averaging method ($k=75$) when quantified in this manner, and better performance from the Gaussian convolution. The boundary bias is extreme
in both cases as the kernels attempt to retrieve information where none should exist. The effect is stronger for the local averaging method because
its zeroth-order kernel does not decay at the tails. The nodes in the middle of the domain do not suffer to the same extent because the individual
kernels have decayed sufficiently prior to reaching the boundary.

Three steps are applied to correct the boundary errors: $k$-nearest neighbour (\textit{knn}) windowing, volume correction, and automatic adaptivity of
the Gaussian kernel parameter. The \textit{knn} window function truncates the kernel to a boundary defined by the $k^{th}$ nearest neighbour. The
volume correction compensates for a kernel that is truncated at the edge of the convex hull of points. The kernel parameter adaptivity adjusts the
kernel parameter according to how centered the query point is within its group of neighbours. The final effect of the boundary corrections is pictured
in \autoref{Fig:AdjustedComparison}, where each of the analyzed kernels shown positioned on the domain boundary. While both the rectangular kernel and
uncorrected Gaussian attempt to take information from outside the domain -- information which does not exist -- the corrected kernel is truncated at
the boundary and changes its volume and width in a corresponding manner.

Limiting the integral approximation to only the nearest $k$ neighbours means $x_a$ need not be centered in $\Omega$ anymore, as long as it is near the
center of the hypercube enclosing itself and its neighbours, $\Omega_{a,k}$. It also saves the computational cost of operating over the entire
dataset. Since neighbouring is a practical necessity when operating over unstructured data, finding the first $k$ neighbours should not be a
significant additional computational burden.

A volume correction factor can then be applied,
	\begin{gather} \label{eqn:ErrorFcnScaling:RhoPrime}
	  \hat{\rho}'_a = \hat{\rho}_a  \frac{V_{\mathbb{R}^d}}{V_{\Omega}},
	\end{gather}
where
	\begin{gather} \label{eqn:ErrorFcnScaling:VolRatio}
	  \frac{V_{\Omega}}{V_{\mathbb{R}^d}} = \left( \frac{1}{2} \right)^d \prod_{k = 1}^{d} \left( \text{erf}\left(\frac{\min |x_{a,d}-\partial \Omega_{a,k,d}|}{\sigma_d / \sqrt{2}}\right)  +  1 \right).
	\end{gather}
This recognizes that the convolution does not truly extend over $\mathbb{R}^d$, but $\Omega_{a,k}$. The previous normalization factor
\eqref{eqn:Norm0} is no longer appropriate. The volume correction factor approximately scales the density estimate to the correct value. The argument
to the error function in \eqref{eqn:ErrorFcnScaling:VolRatio} is the minimum distance from $x_a$ to the boundary along each dimension $d$, $\partial
\Omega_{a,k,d}$, scaled by $\sigma_d$. Truncation is only performed on the nearest boundary in each dimension in equation
\eqref{eqn:ErrorFcnScaling:VolRatio} because numerical results indicate a slight underestimate in point density is more beneficial than an
overestimate. The total effect is to essentially rescale the convolution so it would integrate to unity on the reduced domain.

The final correction is to vary $\sigma$ based on the proximity of $x_a$ to $\partial \Omega_{a,k}$. It is optimal for the Gaussian kernel to decay
sufficiently by the time it reaches the truncated portion of the domain, but without becoming so highly localized that the nodes chosen to be included
do not contribute significantly to the estimate. Using
	\begin{gather} \label{eqn:AutoKernel}
	  \sigma_i = \max | x_{a,i} - \partial \Omega_{a,i} | / N_\sigma,
	\end{gather}
with $N_\sigma \approx 3$ achieves an appropriate amount of decay. An interesting alternative to this approach would be to determine principal
directions of the node distribution via PCA or SVD and select $\sigma$ along those directions in the manner of equation \eqref{eqn:AutoKernel}. The
resulting $\Sigma$ would generally not be diagonal after rotation. This option has not been considered here.

Finally, the nodal spacing may be approximated by inserting $\hat{\rho}'$ in place of $\rho(x)$ in equation \eqref{eqn:ha},
	\begin{gather*}
	  h_a = \frac{1}{\hat{\rho}'(x_a)^{1/d}}
	\end{gather*}
The result of each successive correction is seen in \autoref{Fig:AdjustedComparison} by comparison to nodes on $[0,1]$ with spacing of $h_a=0.01$.
After all corrections are applied, the maximum error in this example is less than 1.5\% on the boundary of the domain. Locations $10$ nodes from the
boundary are accurate to within $0.1$\% of the true value.

The final algorithm for $h_a$ is given by Algorithm \ref{alg:ha}.
		\begin{algorithm}[h!]
		\caption{Adaptive nodal spacing parameter, $h_a$}
			\begin{algorithmic}[1]
				\For{$x_a$, $a = 1,\ldots,N_a$}
					\State Find $k$ nearest neighbours and determine the volume and boundaries of $\Omega_{a,k}$
					\State Solve for components of $\Sigma$ using equation \eqref{eqn:AutoKernel}.
					\State Evaluate $\hat{\rho}(x_a)$ with equation \eqref{eqn:DensityApprox}.
					\State Evaluate $\hat{\rho}'$ with equation \eqref{eqn:ErrorFcnScaling:RhoPrime}.
					\State Evaluate $h_a$ using equation \eqref{eqn:ha}.
				\EndFor
			\end{algorithmic}
		\label{alg:ha}
		\end{algorithm}

\begin{figure}[h!]
			\centering
			\includegraphics[width=0.48\textwidth]{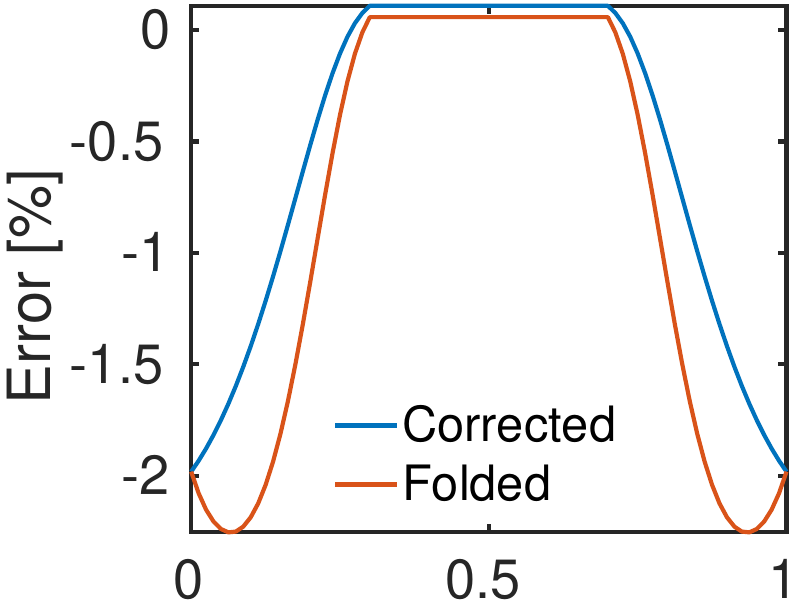}
			\caption{Error in the estimation of $h_a$ in the interval $[0,1]$ for a homogeneous distribution of size $h_a=0.01$ as shown in \autoref{Fig:ErfInfluence}. Proposed correction method \textit{vs.} folded Gaussian \cite{Rice1984,Hall1991,Marron1994}  kernel density estimation obtained with $k=50$ nearest neighbors and using a $\sigma$ scaling factor of $N_\sigma=2.9$ in the example. }
			\label{Fig:Folded_vs_Normal}
		\end{figure}

It should be noted that this approach to correcting boundary bias does not align perfectly with the standard KDE methodology. While using a window
function to truncate the Gaussian kernel is common, the other two steps are variations chosen to operate sufficiently and conveniently without
supervision. The volume correction factor resembles a folded distribution which is common in KDE \cite{Rice1984,Hall1991,Marron1994}. Experiments
using the same error quantification as \autoref{Fig:haPerformance1d} revealed the folded normal distribution to perform similarly or slightly worse
than the volume corrected normal distribution for this application (\autoref{Fig:Folded_vs_Normal}). Since there is extra complexity involved in using
a folded distribution in multiple dimensions \cite{Chakraborty2013}, the simple volume correction is used instead. Similarly, algorithms have been
developed for choosing kernel parameters -- a variety of which are described in \cite{Silverman1998} -- but these are often unnecessarily involved for
a KDE scheme meant to opperate cleanly as a background task. A similar argument can be made for using skewed distributions. They would likely be
effective, but the extra difficulties involved in setting their parameters is unnecessary for this application. The simple boundary corrections
proposed here are suitable for the context of this work.

		\subsubsection{HOLMES Accuracy with Adaptive $h_a$}
		\label{subsubsec:haAcc}
Equipped with an adjusted HOLMES formulation and a method for varying the nodal spacing accurately, the effect of an adjustable kernel width on HOLMES
interpolation accuracy can now be assessed.

Rastrigin's function (\ref{app:TestFunc}) is used for this analysis. The error in a HOLMES interpolation is quantified via the discrete $\ell_2$ norm,
\begin{gather} \label{eqn:l2Norm}
  \ell_2 = \left[\sum_{p=1}^{N_p} \left(u(x_p) - u_I(x_p) \right)^2 / N_p \right]^{1/2}.
\end{gather}
In these tests $x_p$ is a grid of $100 \times 100$ grid points covering the domain $[0,1]^2$, which was found to produce consistent results. HOLMES
functions were evaluated with parameters $p=3$, order $3$, and $\gamma=0.12$.

Tests compared the performance of HOLMES when the kernel width was adapted in three separate ways: the KDE method proposed here using a modified
Gaussian kernel, a simple KDE method using a rectangular kernel (an average of local values), and a non-adapted kernel where the global average of
node spacing is used for all nodes. In all cases HOLMES was confirmed to converge at the appropriate rate as determined by its order of polynomial
consistency.

\begin{figure}[h!]
		\centering
		\begin{subfigure}[t]{\textwidth}
		  	\centering
			\includegraphics[width=0.66\textwidth]{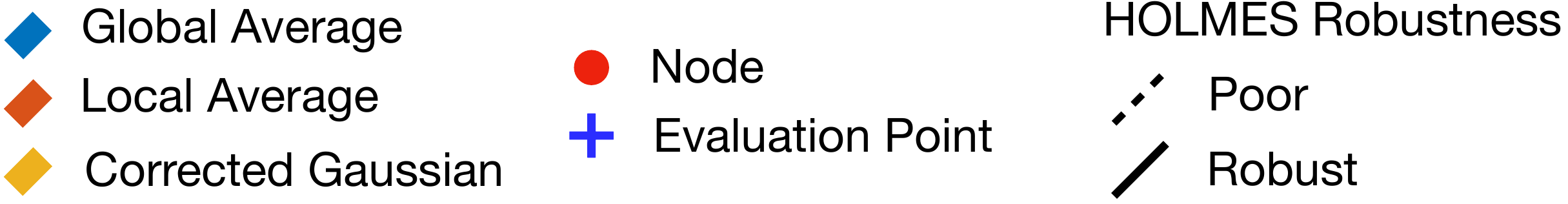}
		\end{subfigure}
		\begin{subfigure}[b][][t]{0.48\textwidth}
			\centering
			\includegraphics[width=\textwidth]{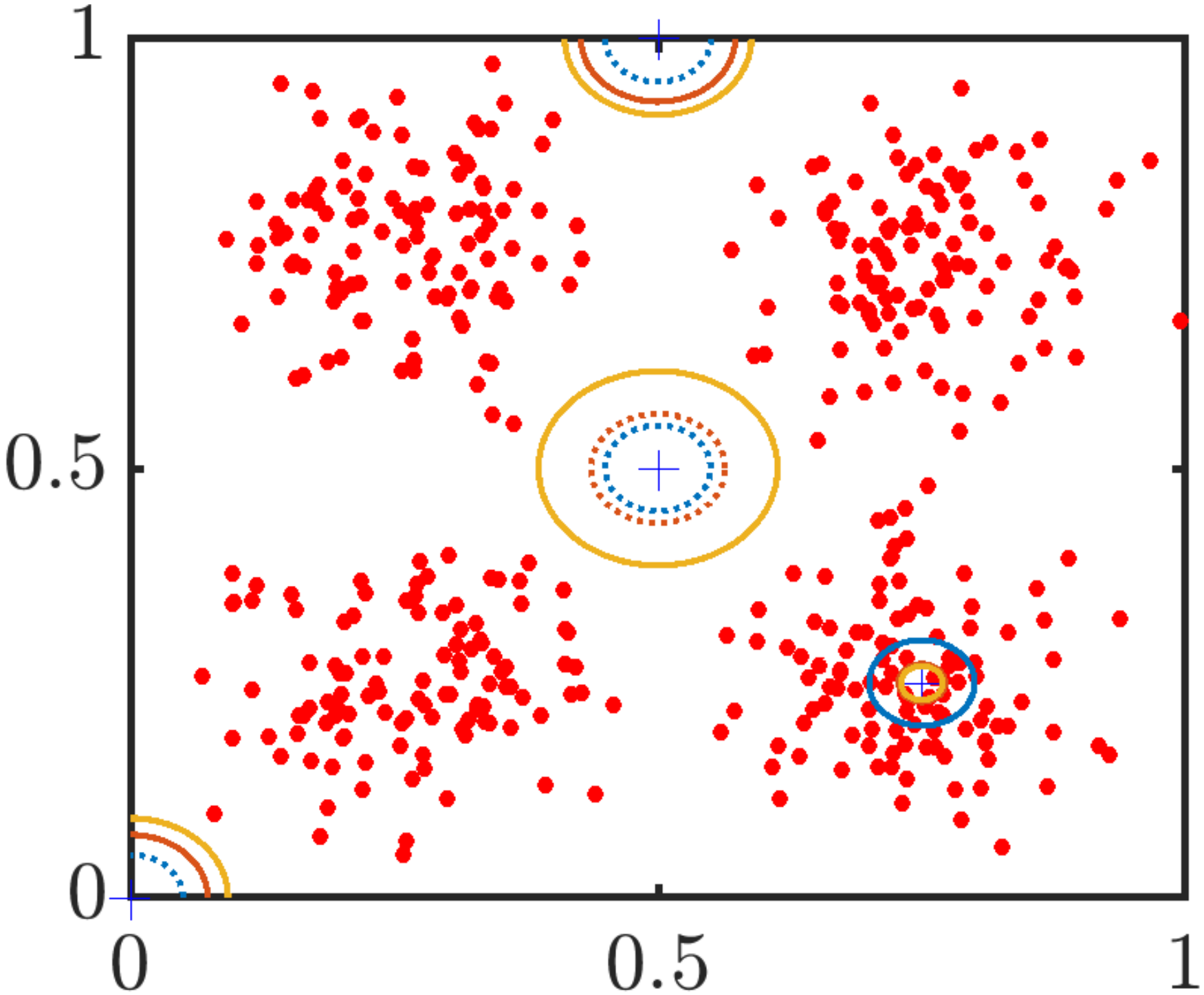}
			\caption{Node distribution with four clumping centers.}
			\label{Fig:ClumpingCompRad}
		\end{subfigure}
		\hfill
		\begin{subfigure}[b][][t]{0.50\textwidth}
			\centering
			\includegraphics[width=\textwidth]{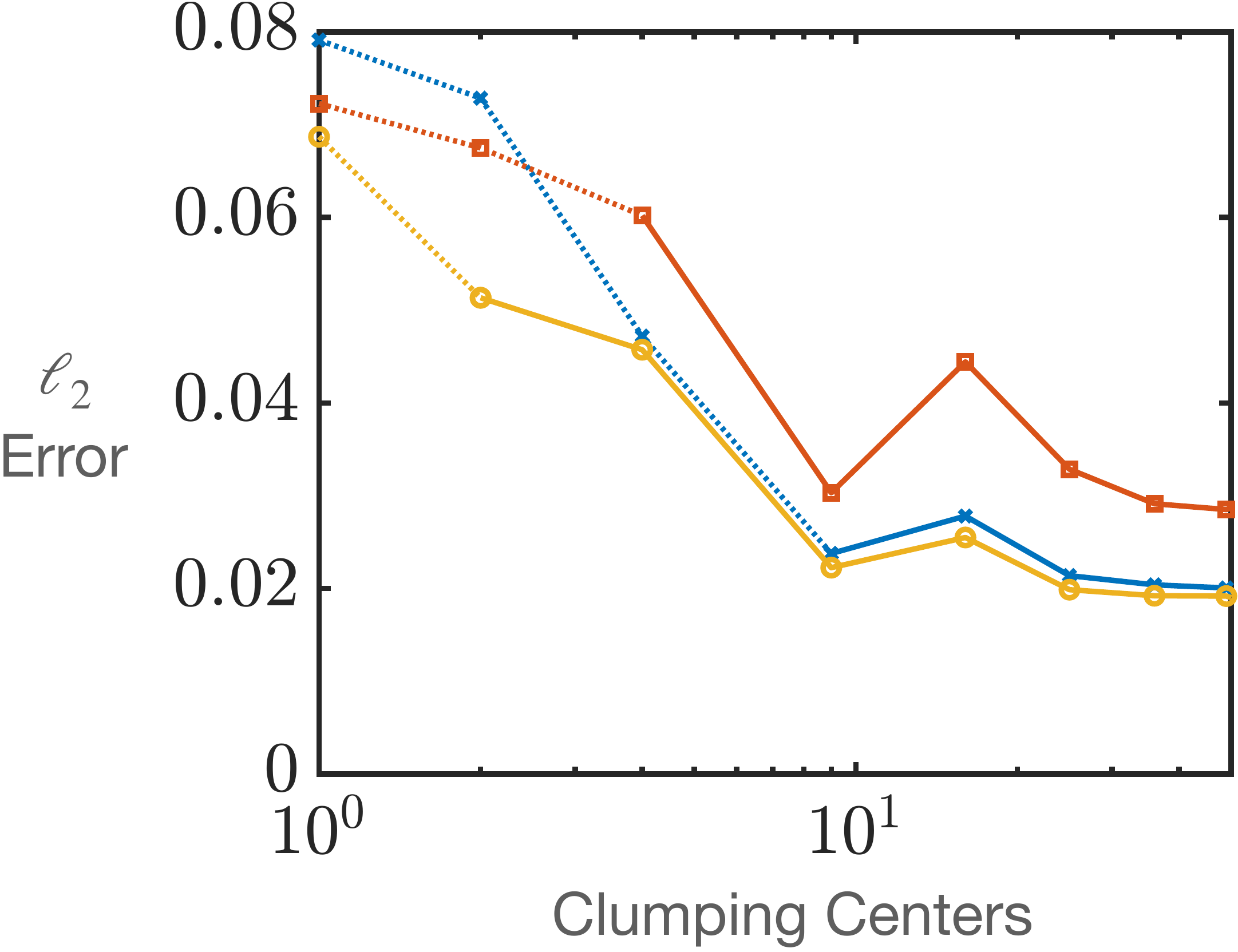}
			\caption{$\ell_2$ error of Rastrigin's Function using HOLMES.}
			\label{Fig:ClumpingCompL2}
		\end{subfigure}
		\caption{Performance comparison of the global average, local average (rectangular kernel)  \eqref{eqn:ha}, and boundary corrected
		  Gaussian \eqref{eqn:dens_approx_Gauss} methods for assigning the spacing parameter, $h(x_a)$. (a)
		  Spacing parameter estimated at different locations  (indicated by a $+$ marker) for a four cluster clumped data. (b) $\ell_2$ norm
		  error evolution as a function of the clumping centers for Rastrigin's function evaluated with the three different spacing parameter
		methods on a $100 \times 100$ grid points covering the domain $[0,1]^2$. Continuous and dashed lines indicate robust and poor
	      performance of the HOLMES metamodel using  $p=3$, order 3, and $\gamma=0.12$.}
		\label{Fig:ClumpingComp}
	\end{figure}

To assess the performance of HOLMES and the $h$-approximation schemes on unevenly distributed, or \textit{clumped} data the $\ell_2$ error can be
calculated over datasets with varying degrees of clumping. Clumping is quantified here by specifying a number of clumping centres around which data
points are normally distributed. The standard deviation of the distribution is $1/4$ the distance between clumping centers. Figure
\ref{Fig:ClumpingCompRad} visualizes this. The accuracy of HOLMES can be plotted against the number of clumping centers, as seen in figure
\ref{Fig:ClumpingCompL2}.

The results of \autoref{Fig:ClumpingCompL2} demonstrate a few things. Notably, the proposed adaptive method achieves better accuracy than either of
the alternative approaches, and is more robust. When using a constant value for $h_a$ HOLMES performs well if the nodes are more evenly distributed;
however, accuracy drops below other approaches as the data becomes more uneven.

More concerning is the difficulty evaluating HOLMES in the highly clumped region using a constant global value for $h_a$. If the convex optimization
problem required to determine the Lagrange multipliers can be solved within a reasonable number of Newton algorithm iterations -- generally less than
20 -- it is termed \textit{robust}. If Newton's method fails to converge, or takes an exceedingly long time to do so, such as more than $1000$
iterations, it is not robust (and termed \textit{poor}).

The rectangular kernel approach to adapting $h_a$ is interesting in that it produces a less accurate HOLMES interpolation in general compared to both
the proposed KDE method and the constant $h$ method; however, it does generally remain fairly robust in the highly clumped tests, as desired.

The proposed KDE approach seems to perform the best with both even and uneven datasets, offering a good balance of robustness and accuracy compared to
the others. No method was able to be completely robust at the extremes of clumping with the parameters used in this test, as seen on the far left
side of \autoref{Fig:ClumpingCompL2}, where all methods have a dotted line, representing poor HOLMES robustness.

	\subsection{Noise Resistance}
	\label{subsec:HOLMESNoise}

A probability distribution which maximizes information entropy will have the greater resistance to errors caused by Gaussian white noise, which itself
is the highest entropy form of noise \cite{Shannon1948}. HOLMES shape functions cannot be viewed as probability distributions because they lack a
non-negativity constraint -- a constraint necessarily lost in order to achieve higher orders of polynomial consistency
\cite{Bompadre2012,Cyron2009}; however, their formulation still depends on maximization of entropy, so they should exhibit resistance to
Gaussian noise compared to most other shape functions.

Radial basis function interpolation is chosen to contrast HOLMES interpolation because of its huge popularity and use in other DoE applications. It is
one of a few other methods which are effective at interpolating Hessian information over unstructured multidimensional data; however, in the forms
analyzed, RBF interpolation performs worse than HOLMES when the data contains noise.

	\begin{align}
		f_N(x) &= f(x)\left( 1 + \zeta G \right) \label{eqn:NoiseStat} \\
		f_N(x) &= f(x) + \zeta\max_{x\in[0,1]^2}(f(x)) G \label{eqn:NoiseNotStat}
	\end{align}

HOLMES and RBF interpolation are compared in two related, but distinct scenarios which an experimentalist may encounter: pure white Gaussian noise,
which injects normally distributed noise into the data, and proportional Gaussian noise, which depends on the amplitude of the underlying signal.
These are represented by \eqref{eqn:NoiseStat} and \eqref{eqn:NoiseNotStat} respectively. Here $G$ is a Gaussian noise generator with a mean of 0 and
standard deviation of 1, and $\zeta$ scales the amplitude of the noise. The first case of constant variance noise could occur when measurements are
limited purely by the accuracy of the experimental apparatus. The second case could arise when experimental physics incite additional errors.

In both scenarios a standard convergence analysis on a regular grid can be carried out using a known function -- Branin's function is used here
(\ref{app:TestFunc}) -- and the convergence of HOLMES and RBF interpolations compared as the noise magnitude is adjusted. In all tests the $\ell_2$
error is plotted as a function of grid size. The $\ell_2$ norm is evaluated using a $100 \times 100$ grid evenly spaced over the function domain
$\Omega = [0,1]^2$.

	\begin{figure}[h!]
		\centering
		\begin{subfigure}[t]{\textwidth}
		  	\centering
			\includegraphics[width=0.3\textwidth]{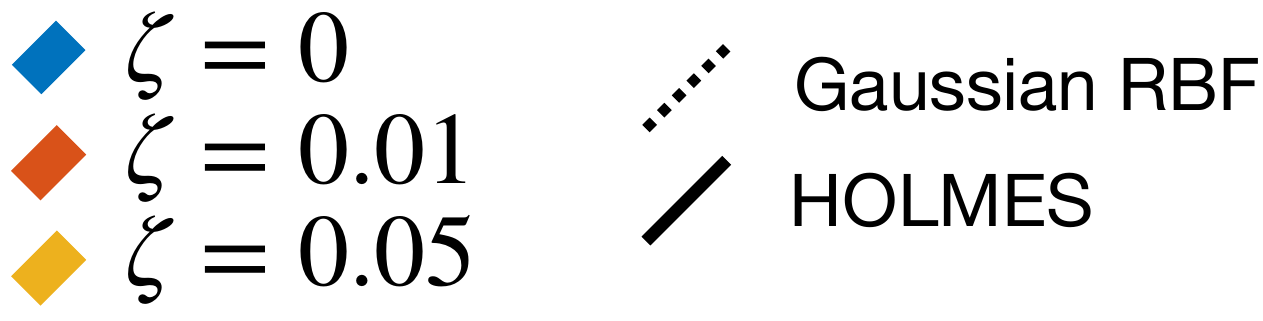}
		\end{subfigure}
		\begin{subfigure}[b][][t]{0.48\textwidth}
			\centering
			\includegraphics[width=\textwidth]{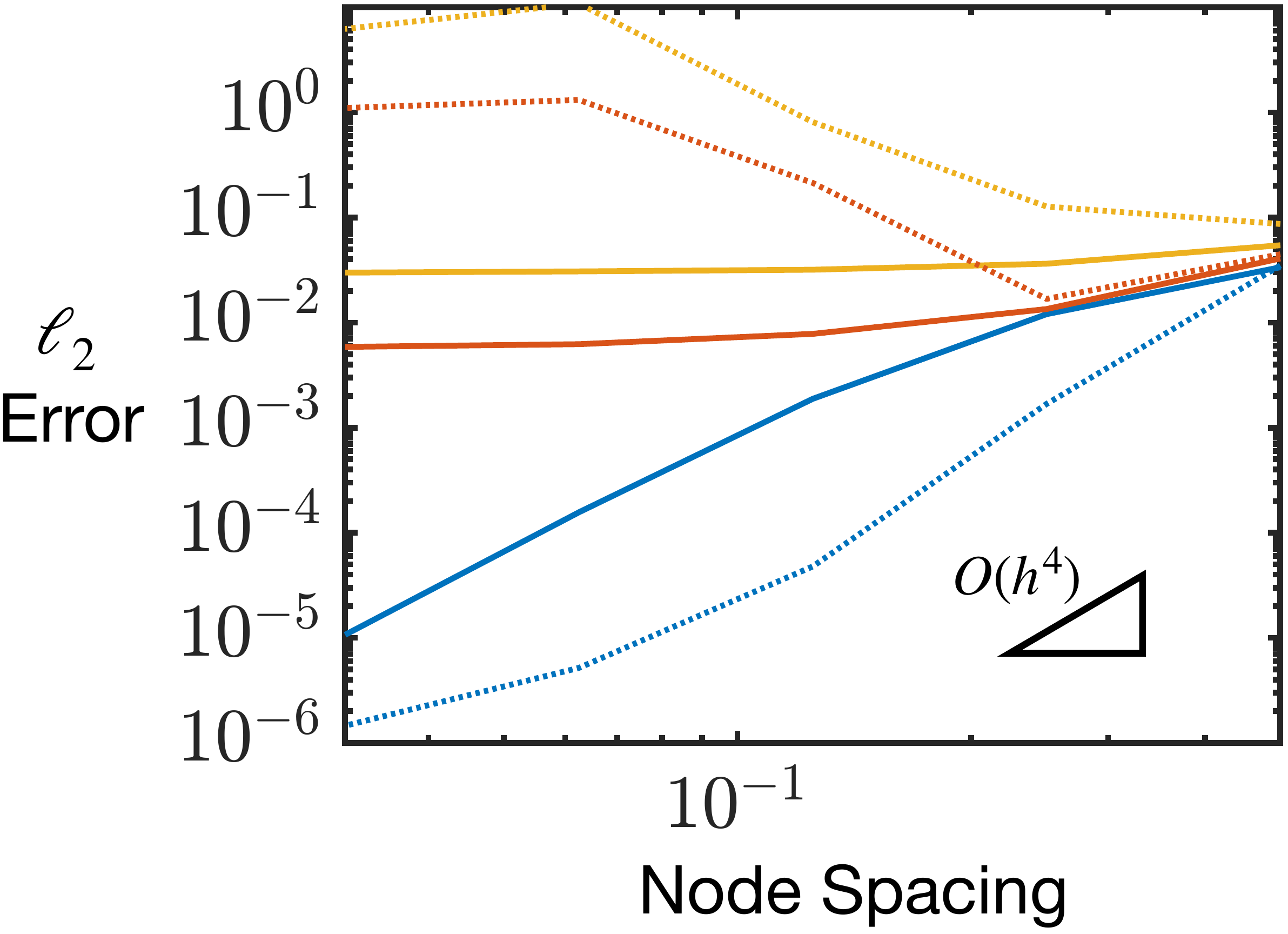}
			\caption{Stationary white noise  \eqref{eqn:NoiseStat}.}
			\label{Fig:NoiseRBFGaussian_Stat}
		\end{subfigure}
		\hfill
		\begin{subfigure}[b][][t]{0.48\textwidth}
			\centering
			\includegraphics[width=\textwidth]{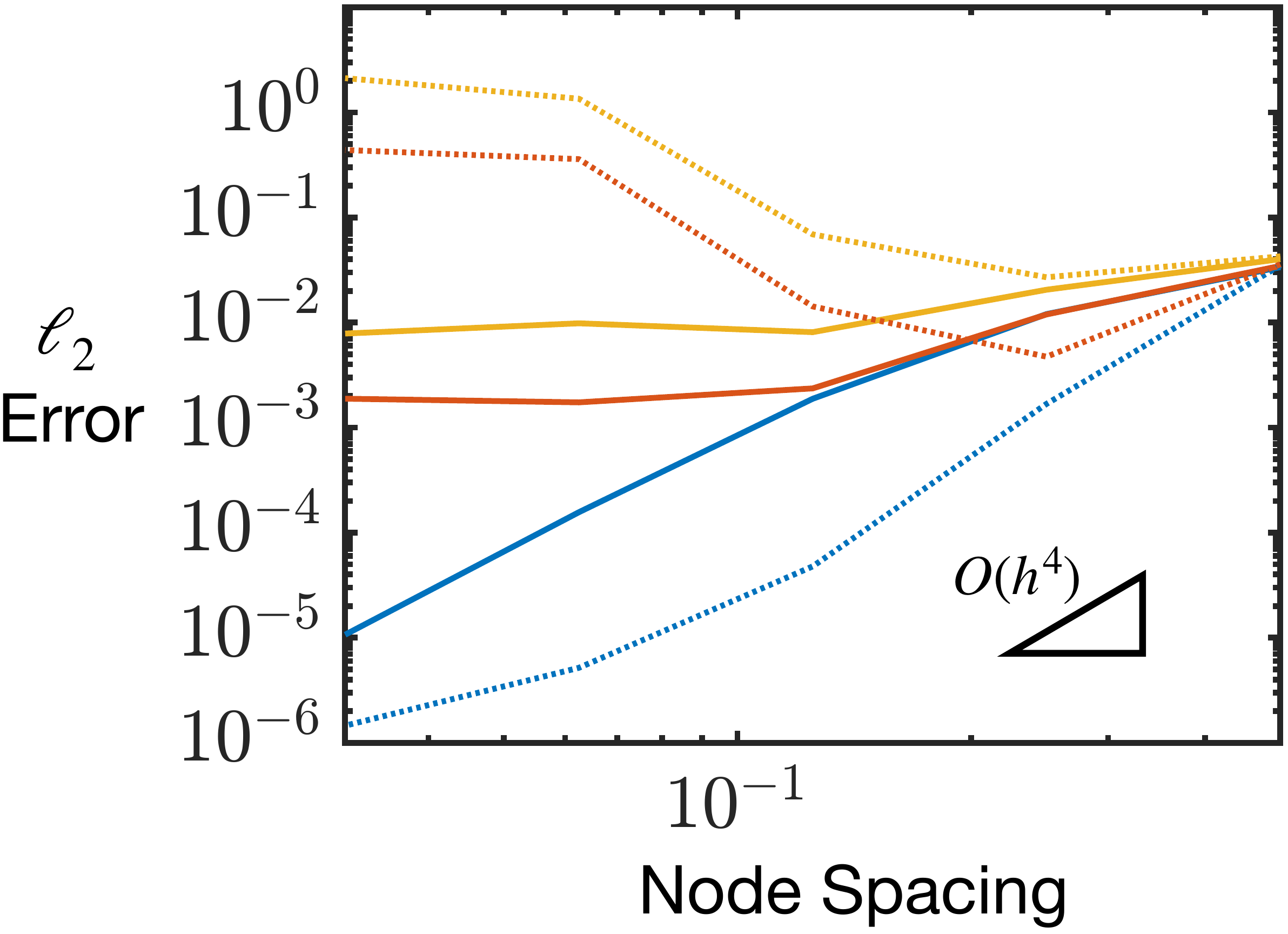}
			\caption{Proportional noise \eqref{eqn:NoiseNotStat}.}
			\label{Fig:NoiseRBFGaussian_Prop}
		\end{subfigure}
		\caption{$\ell_2$ error for a Gaussian hill function with varying noise amplitude as given by \eqref{eqn:NoiseStat} and
		\eqref{eqn:NoiseNotStat} using RBF-Gaussian interpolation (dotted lines) and the HOLMES approximants (solid lines). $\mathcal{O}(h^4)$
		scaling shown with dot-dashed line. $\zeta$ correspond to the white noise amplitude used. HOLMES approximants computed using $\gamma
		\approx 0.8$ for a distance norm of $p=2$.}
		\label{Fig:NoiseRBFGaussian}
	\end{figure}

Possibly the most popular RBF kernel is the Gaussian. \autoref{Fig:NoiseRBFGaussian} compares the influence of noise on both RBF-Gaussian and HOLMES
interpolation schemes, represented by dotted and solid lines respectively. The Gaussian RBF significantly outperforms HOLMES in the noiseless case of
$\zeta=0$ due to it being an exact interpolant rather than an approximant. Despite this, the addition of even 1\% noise can have devastating effects
on RBF accuracy. The convergence plots do not plateau at some level proportional to the noise magnitude, but explode and entirely ruin the
interpolation. RBF-Gaussian is known to be highly sensitive to its kernel parameter \cite{Buhmann2003}, but even extensive manual manipulation could
not achieve better results than HOLMES for this case, making it a poor choice for an interpolation scheme in the presence of noise.

	\begin{figure}[h!]
		\centering
		\begin{subfigure}[t]{\textwidth}
		  	\centering
			\includegraphics[width=0.30\textwidth]{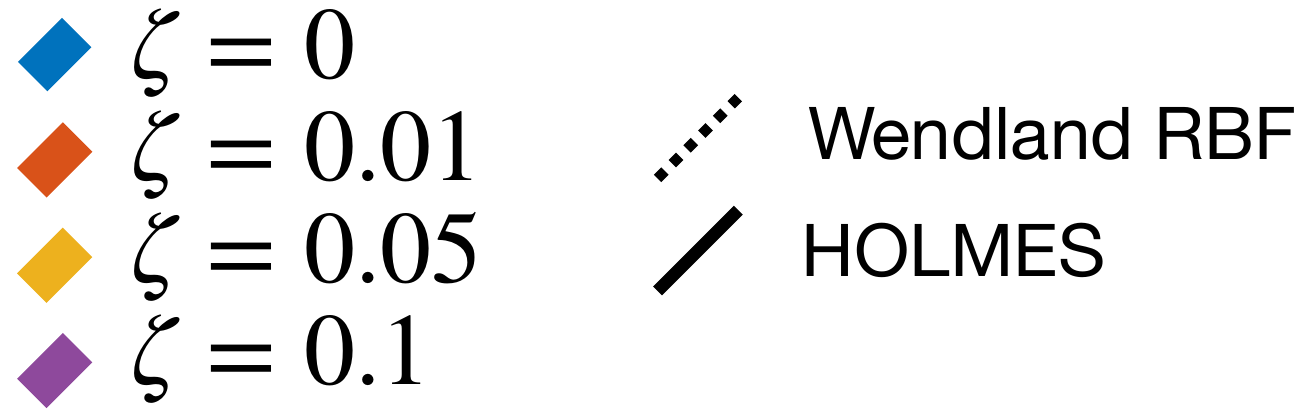}
		\end{subfigure}
		\begin{subfigure}[b][][t]{0.48\textwidth}
			\centering
			\includegraphics[width=\textwidth]{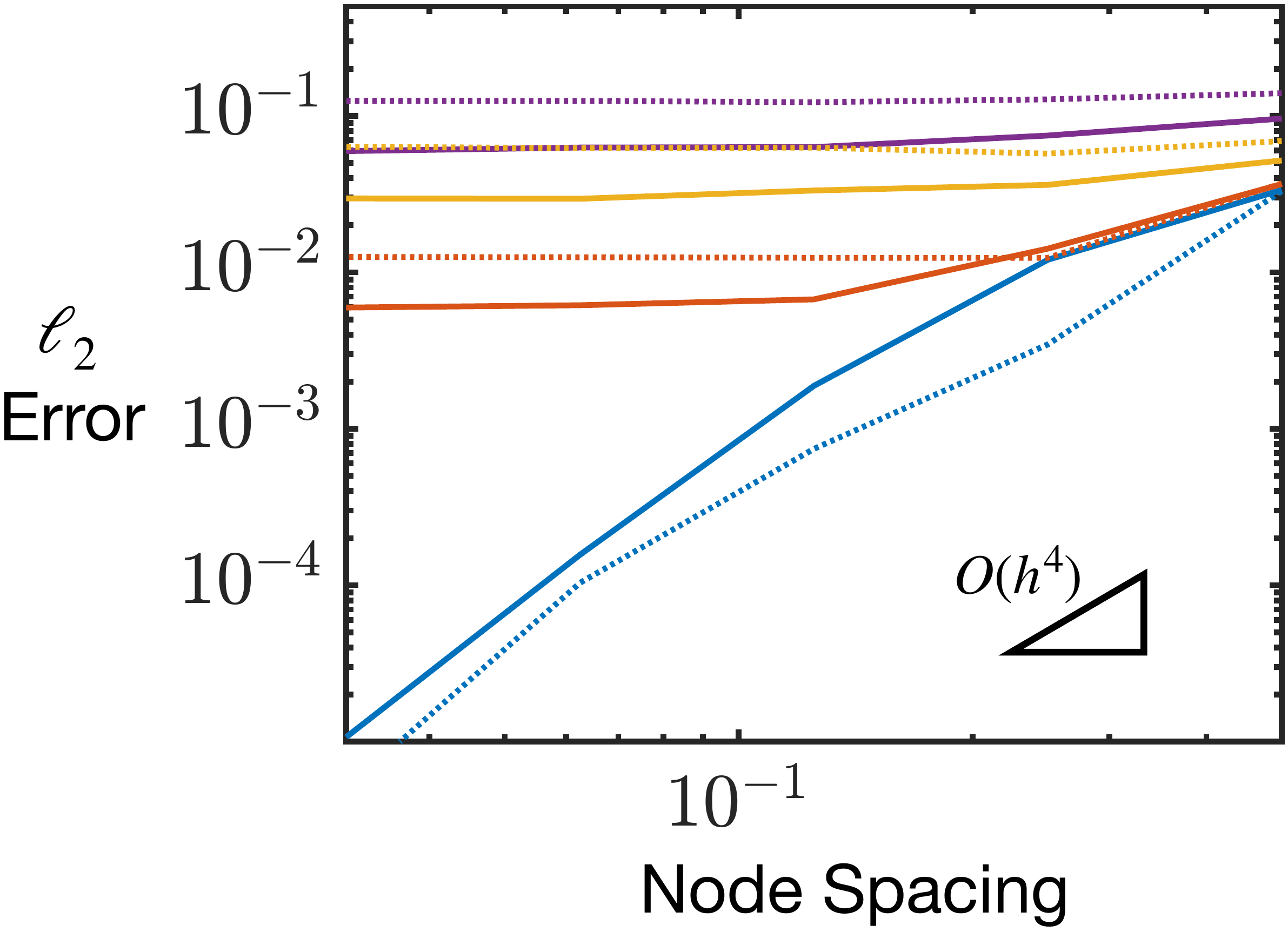}
			\caption{Stationary white noise  \eqref{eqn:NoiseStat}.}
			\label{Fig:NoiseRBFWendlandStat}
		\end{subfigure}
		\hfill
		\begin{subfigure}[b][][t]{0.48\textwidth}
			\centering
			\includegraphics[width=\textwidth]{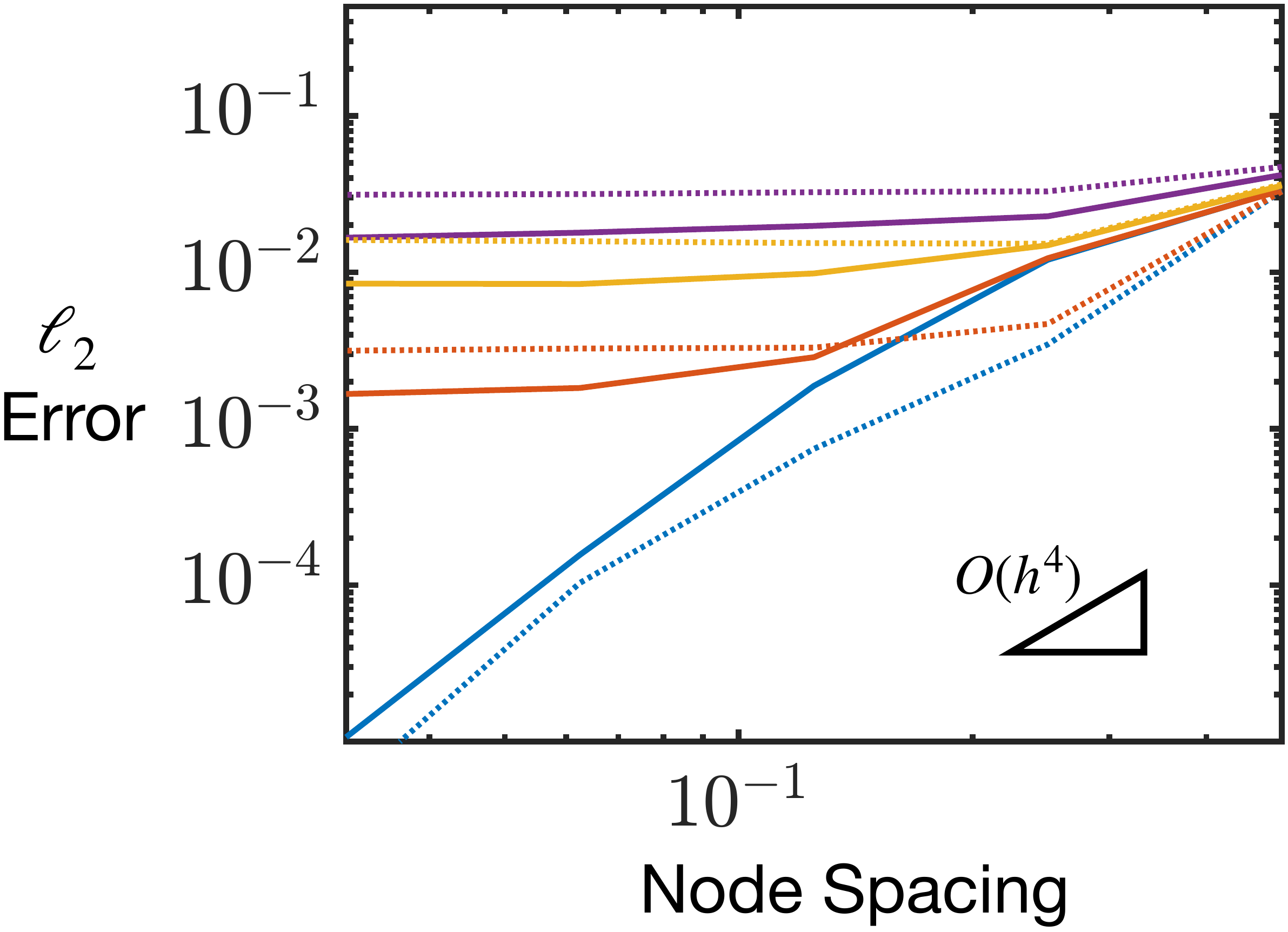}
			\caption{Proportional noise \eqref{eqn:NoiseNotStat}.}
			\label{Fig:NoiseRBFWendlandProp}
		\end{subfigure}
		\caption{$\ell_2$ error for a Gaussian hill function with varying noise amplitude as given by \eqref{eqn:NoiseStat} and
		\eqref{eqn:NoiseNotStat} using RBF-Wendland Interpolations (dotted lines) and the HOLMES approximants (solid lines).
		$\mathcal{O}(h^4)$ scaling shown with dot-dashed line. $\sigma$ correspond to the white noise amplitude used. HOLMES approximants
		computed using $\gamma \approx 0.8$ for a distance norm of $p=2$.}
		\label{Fig:NoiseRBFWendland}
	\end{figure}

An analysis of RBF performance would be amiss to neglect the popular family of Wendland kernels \cite{Wendland2005}, whose compact support make them
more computationally practical than the Gaussian kernel. \autoref{Fig:NoiseRBFWendland} summarizes this analysis, where vastly superior noise
resistance is found compared to the RBF-Gaussian interpolation, but accuracy is still inferior to HOLMES interpolation. HOLMES and RBF-Wendland are
using identical cutoff radii in this analysis to make a more equal comparison, but the behaviour with noise is similar regardless of how large a
cutoff radius RBF-Wendland is allowed. RBF-Wendland consistently plateaued at around 150\% the error of HOLMES once noise was introduced. It should be
noted that RBF-Wendland again performs better than HOLMES in the noiseless case because it is an exact interpolant, but this is the very reason it
performs poorly with noise. RBF interpolation risks overfitting as it attempts to exactly replicate the known data points, whereas HOLMES relaxes this
constraint in favour of maximizing entropy.

On the other hand, HOLMES is relatively insensitive to its locality parameter $\gamma$, which can generally be left at $\gamma \approx 0.8$ for a
distance norm of $p=2$ \cite{Bompadre2012}. Since higher order distance norms cause faster shape function decay, and thus, fewer neighbours, this work
has adjusted $\gamma$ to maintain the same number regardless of norm used through $\gamma(p) = \gamma_0^{p/p_0} (-1-\log(\epsilon))^{1-p/p_0}$, with
$\gamma_0 = 0.8,~p_0 = 2,~\epsilon=2\text{e--}16$. This scales $\gamma$ to maintain the same neighbourhood, and takes $\gamma=0.8$ and $p_0 = 2$ as a
good standard.

RBF interpolants can be made more resistant to data noise making them approximants like HOLMES \cite{Theunissen2018}. This approach is effective in
many situations, but requires some indication of how to choose smoothing parameters. For the common case of unreplicated experiments this paper is
concerned with, where few if any of the statistical features of the data are known, there is no reliable approach for parameter selection. If any
approach is to be applied without manual intervention to increase resolution of computer simulations it becomes clear manual smoothing parameters will
be \textit{ad hoc} and unreliable.


\section{Examples}
\label{sec:Examples}

This section deals with a variety of examples to demonstrate the basic functionality of the proposed DoE method and investigate its parameters. A
comparison is made in \autoref{subsec:InfError} to a comparable method which demonstrates the increased resistance to noise of the entire DoE scheme.
Section \ref{subsec:ADoEforTimeFuncs} then investigates a simple approach to apply the DoE technique to time dependent problems, which could have
interesting future applications to the numerical solution of PDEs.

It should be noted that all investigations in this work use a coarse FF grid as the starting point for the adaptive DoE method to work from. In
general, any spacing filling approach would be appropriate for the initial seeding of data points, with the the user ultimately selecting the method
most appropriate to their specific problem. While it is evident that the initial seed will influence the result of the DoE method, the immense number
of different seeding methods coupled with their performance on different types of problems makes a thorough investigation outside the scope of this
work. A FF seed is chosen for these experiments because of its simplicity, not because it achieves the best results on any particular problem.

	\subsection{Canonical Functions}
	\label{subsec:CanonFunc}

A successful DoE method should reproduce certain results on simple canonical functions. Specifically, the sampling of a radially symmetric function
should be radially symmetric, and that of a plane function should be space filling.  Initially, the proposed method could not achieve this because of
non-existant HOLMES derivatives along the domain boundary \cite{Bompadre2012}, and large numerical errors in derivative evaluations very close to the
boundary. A potential solution could use L'H\^opital's rule to evaluate the derivatives along the boundary \cite{Greco2013}, but would be non-trivial
to implement in multiple dimensions. An easily implmemented practical solution is to recognize that derivatives only suffer significant distortion
very close to the boundary, and restrict the DoE method from considering points within this region. This could be problematic when the region of
interest lies along a boundary, but otherwise does not severely handicap the technique.

\begin{figure}[h]
		\centering
		\begin{subfigure}[b]{0.48\textwidth}
			\centering
			\includegraphics[width=\textwidth]{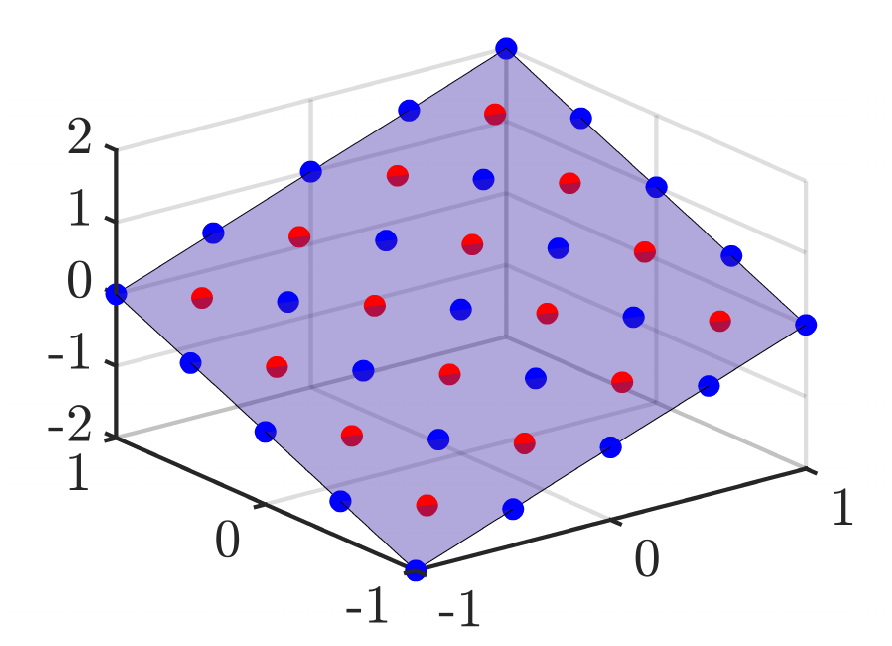}
			\caption{Plane}
			\label{Fig:PlaneDistribution}
		\end{subfigure}
		\hfill
		\begin{subfigure}[b]{0.48\textwidth}
			\centering
			\includegraphics[width=\textwidth]{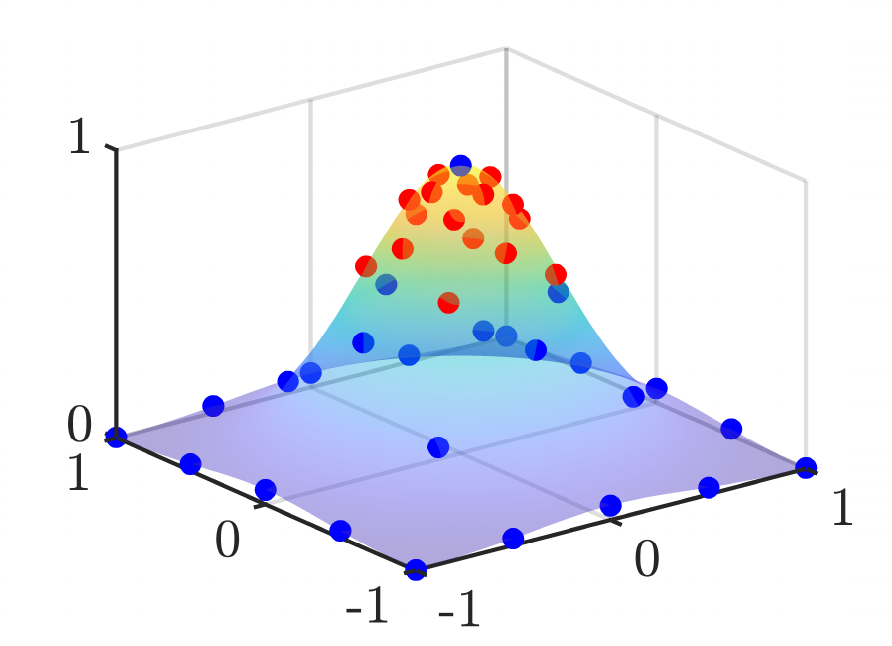}
			\caption{Gaussian Hill}
			\label{Fig:GaussDistribution}
		\end{subfigure}
		\caption{Adaptive DoE nodal distribution for two test functions (a) plane and (b) Gaussian Hill (see \ref{app:TestFunc}). Initially,
		25 points were selected using space filling distribution (red points). Additional 16 points (blue) were placed in the domain according
		to the DoE. The space filling and radial distribution location of points for the respective test functions are as expected. HOLMES
		approximants computed using $p=3$, order 3, and $\gamma = 0.12$.}
		\label{Fig:GoodDistributions}
	\end{figure}

\autoref{Fig:GoodDistributions} demonstrates the approach after suitably restricting the sample point domain. The method now replicates the expected
space-filling and radially symmetric results within the limit of the $41\times41$ evaluation grid.

\begin{figure}[h]
		\centering
		\includegraphics[width=\textwidth]{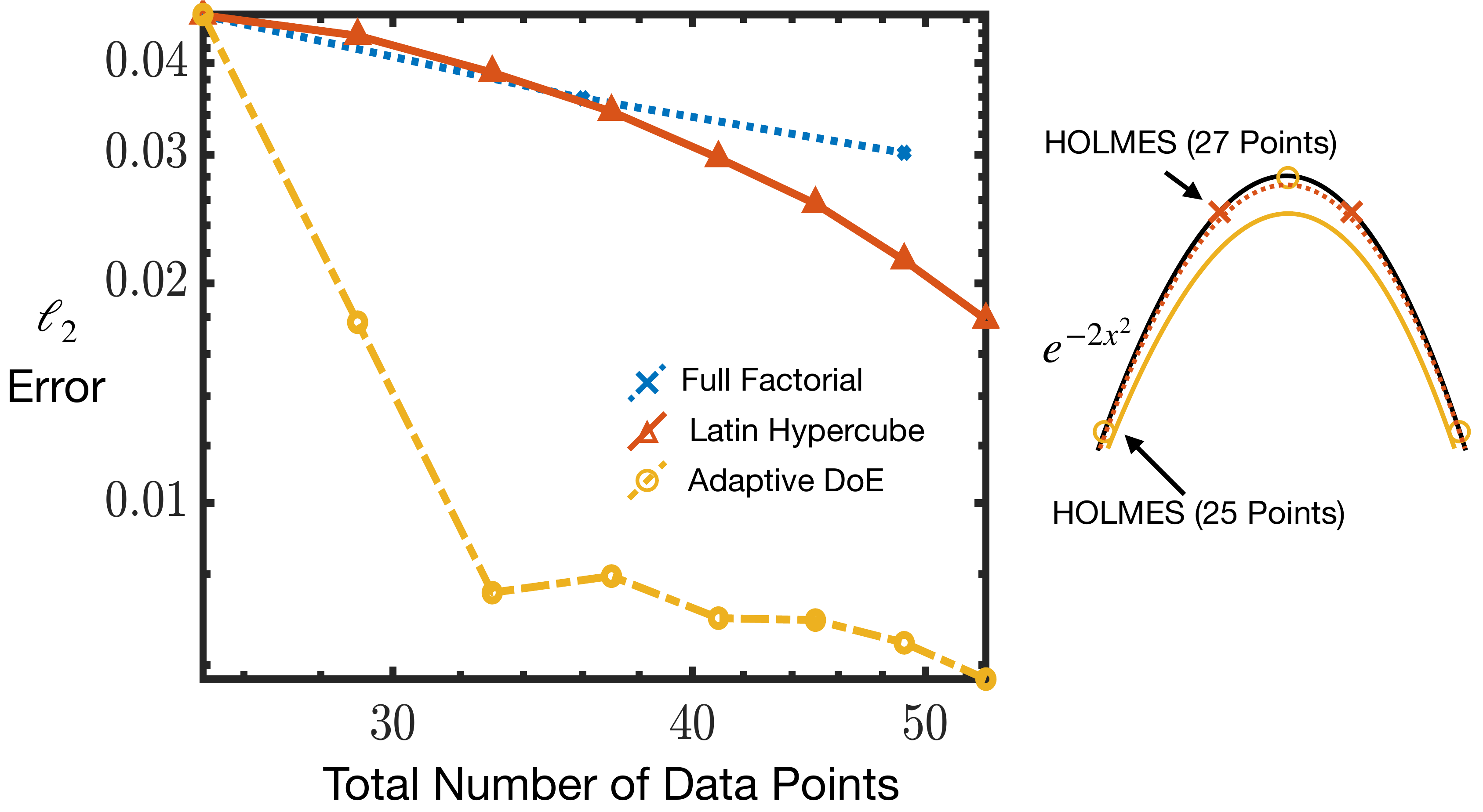} 
		\caption{Comparison of the $\ell_2$ error evolution with the number of data points between the proposed adaptive DoE (dot-dashed
		line), LH (solid line), and FF (dotted line) methods on a Gaussian hill test function (see \ref{app:TestFunc}). The test function was
		evaluated using the metamodels on an evenly distributed $100 \times 100$ grid over the domain $[-1,1]^2$. HOLMES approximants computed
		using $p=3$, order 3, and $\gamma = 0.12$.}
		\label{Fig:GaussConv}
	\end{figure}

Performance of the adaptive method can be compared to the popular non-adaptive methods of LHS and full factorial design by constructing a separate
HOLMES approximation with each distinct set of data points and comparing $\ell_2$ errors over the domain $[-1,1]^2$. The Gaussian hill is used and the
underlying function in this test case, since the plane is perfectly approximated by HOLMES. In this case a $100 \times 100$ evaluation grid was used
to calculate $\ell_2$ error, all methods began with a $5\times5$ grid of data points, and HOLMES parameters of $p=3$ and $\gamma=0.12$ were used.
\autoref{Fig:GaussConv} shows the results. Error is seen to drop rapidly with a few targeted data point additions by the adaptive method, while the
error of non-adaptive methods decreases in a slow, but steady manner. The initial samples are especially impressive, as they reduce the error by a
factor of $5$ with the addition of only $8$ points to the starting set. Evidently, the main source of error occurs at the peak of the Gaussian, and
once that is eliminated the smaller errors distributed throughout the domain will take longer to deal with. This concept is demonstrated on the right
side of \autoref{Fig:GaussConv}, where the strategic addition of two points can massively improve a HOLMES interpolation in 1D. This example is
particularly striking because the errors are concentrated in a small area, and thus quickly eliminated. The benefits of an adaptive DoE technique
would not appear as striking for a function with a less concentrated error profile -- nor would they be expected to be.

It should be noted that throughout this work, $\ell_2$ error results provided by LHS are considered as the average of 100 tests to offset the
intrinsic randomization of the technique. This number of samples provides a reasonable indication of the trajectory of the $\ell_2$ error.

	\subsection{Influence of Kernel Parameter}
	\label{subsec:KernelParam}

\begin{figure}[h]
		\centering
		\begin{subfigure}[b]{0.3\textwidth}
			\centering
			\includegraphics[width=\textwidth]{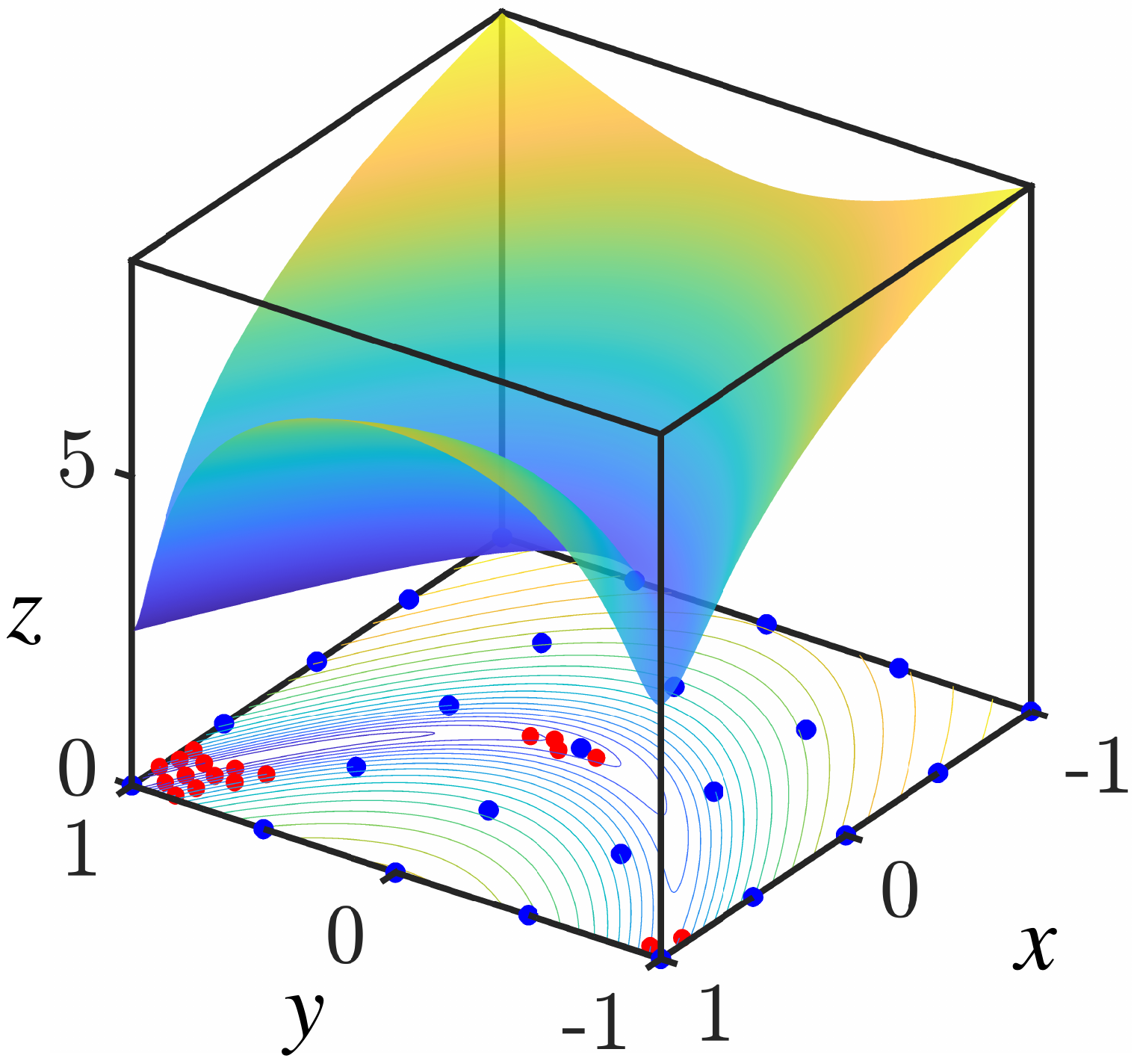}
			\caption{$R_0=0.5$}
			\label{Fig:Rosen05}
		\end{subfigure}
		\hfill
		\begin{subfigure}[b]{0.3\textwidth}
			\centering
			\includegraphics[width=\textwidth]{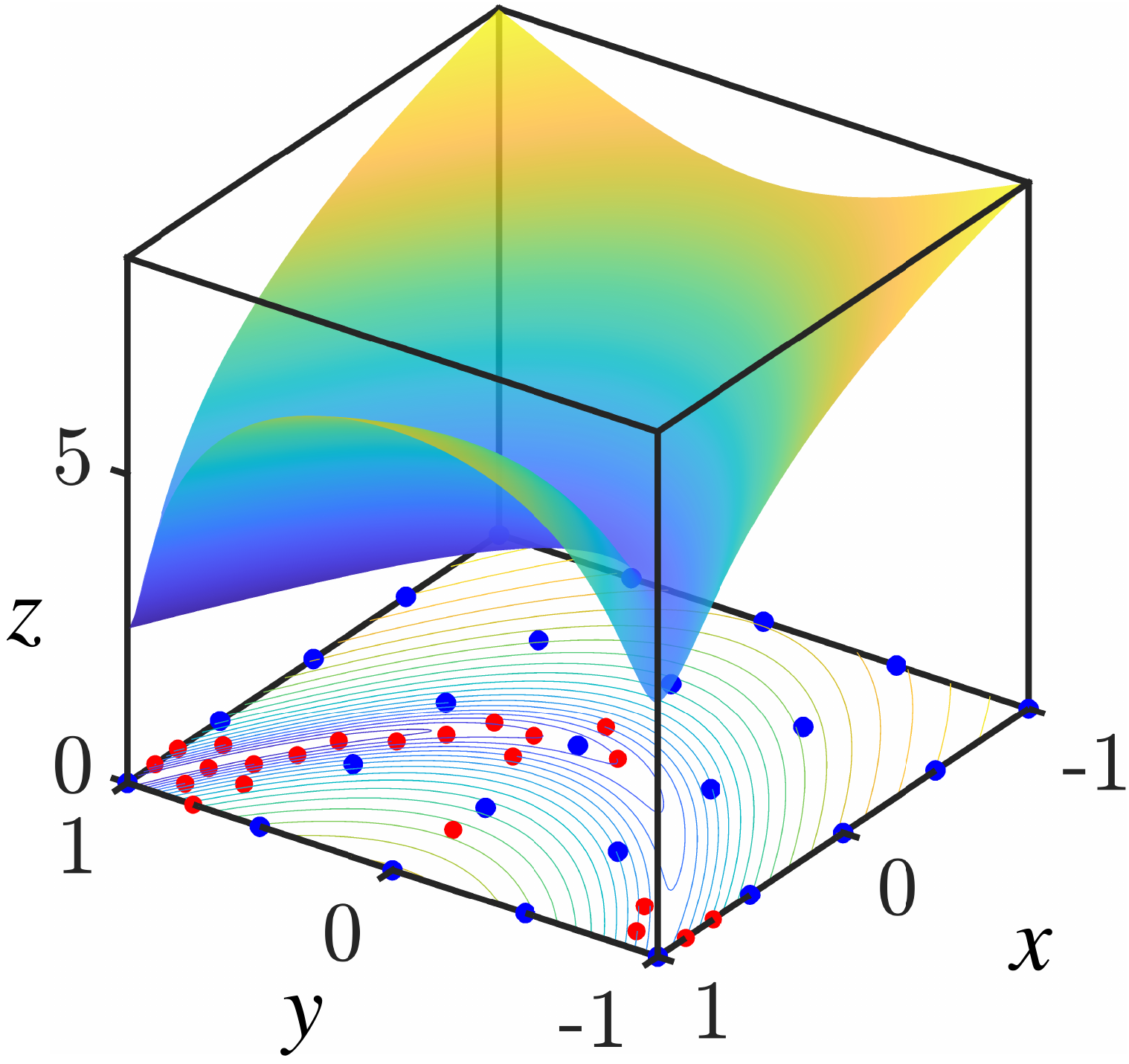}
			\caption{$R_0=1.0$}
			\label{Fig:Rosen10}
		\end{subfigure}
		\hfill
		\begin{subfigure}[b]{0.3\textwidth}
			\centering
			\includegraphics[width=\textwidth]{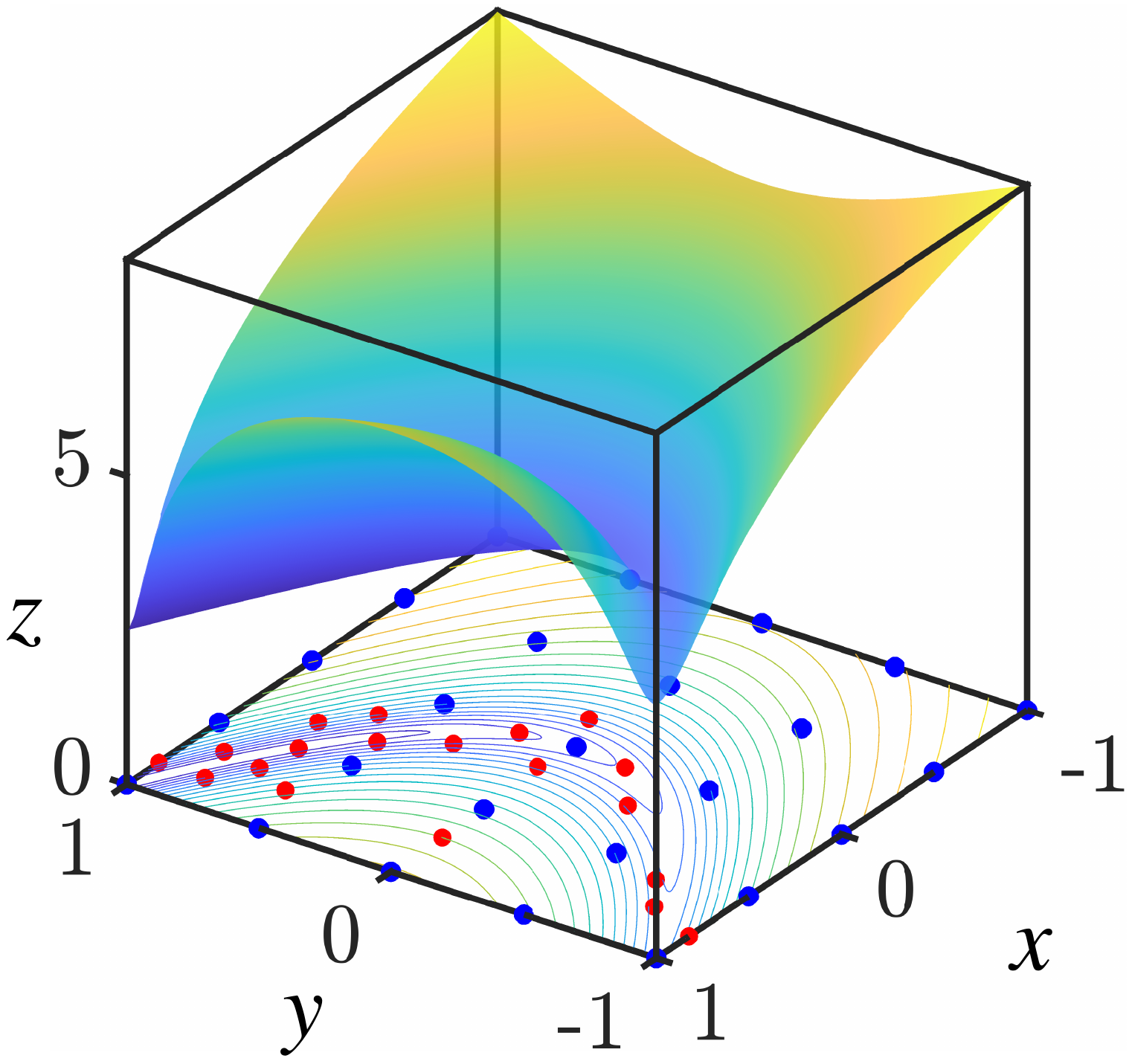}
			\caption{$R_0=1.4$}
			\label{Fig:Rosen14}
		\end{subfigure}
		\caption{Log-scaled Rosenbrock function (see \ref{app:TestFunc}) nodal distributions showing $25$ initial points (blue) evenly
		  distributed in the domain $[-1,1]^2$. Additional $18$ nodes were added using the adaptive DoE technique using different values of
		  the kernel size, $R_0$, to show the influence in the placement of these points. HOLMES approximants computed using $p=3$, order 3,
		and $\gamma=0.12$.}
		\label{Fig:RosenDist}
	\end{figure}

The kernel parameter, $\xi$, used in the spacing function $h(x)$ has a large influence on the resulting point distribution, as mentioned in
\autoref{subsubsec:Q_S}. The choice of scalar, $R_0$, which modifies $\xi$ via \eqref{eqn:kernelfromfill} and \eqref{eqn:suppradfromfill}, has a
significant influence on the DoE method. Figure \ref{Fig:RosenDist} demonstrates this with a log-scaled Rosenbrock function. Smaller values of $R_0$
will allow tighter clumping of points, since $Q_S$ only becomes dominant very close to a data point. An extreme case of this can be found in figure
\ref{Fig:Rosen05}, where the data points are clearly far too concentrated. Larger values will enforce a spacing filling distribution with increasing
severity as $Q_S$ becomes dominant nearly everywhere.

\begin{figure}[h]
		\centering
		\includegraphics[width=.66\textwidth]{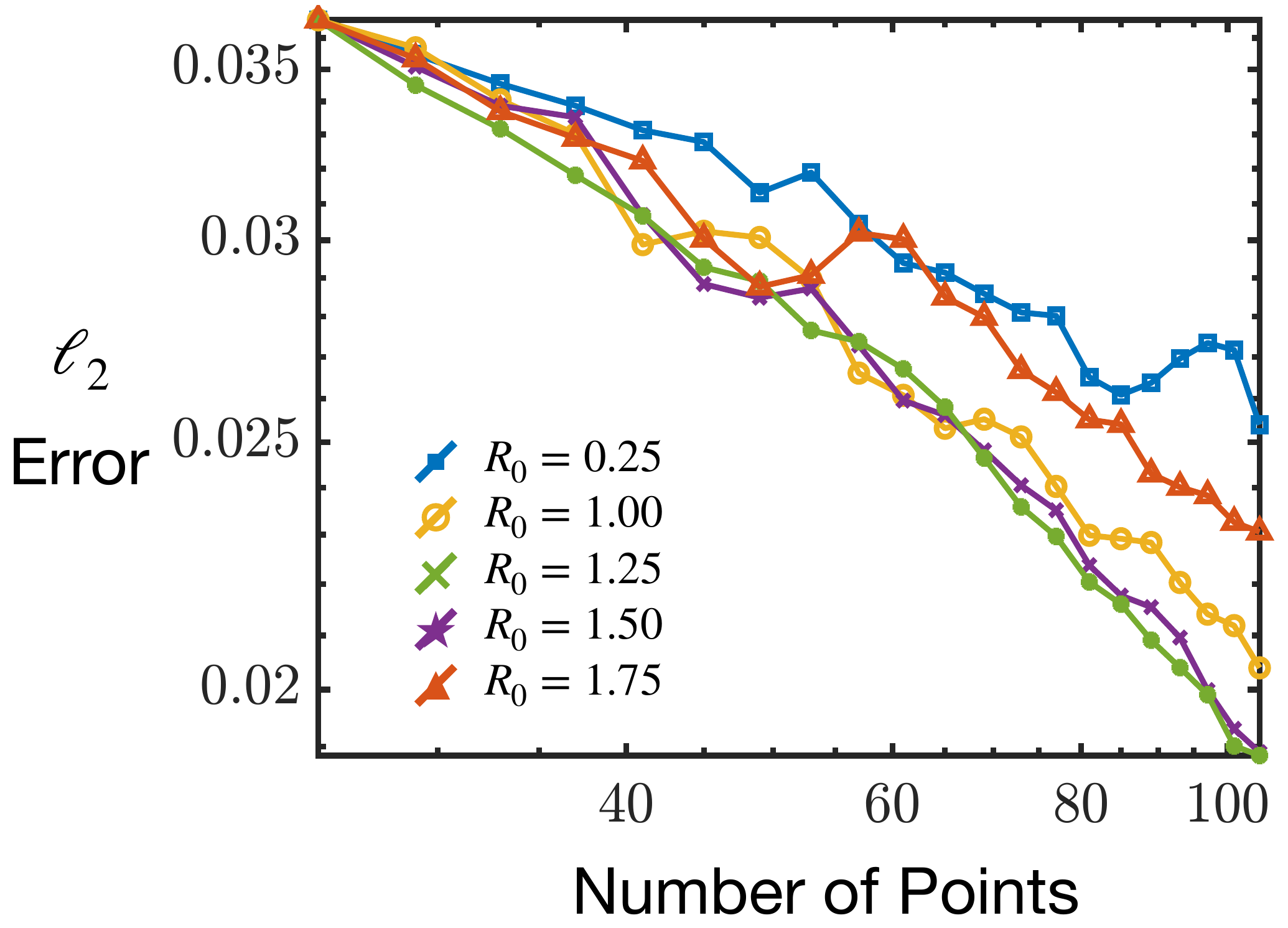}
		\caption{$\ell^2$ error \textit{vs.} number of data points for several values of $R_0=[0.25,1.4]$. The test function used in this test
		  was the log-scaled Rosenbrock function (see \ref{app:TestFunc}) evaluated on an evenly distributed $100 \times 100$ evaluation grid
		over the domain $[-1,1]^2$. HOLMES approximants computed using $p=3$, order 3, and $\gamma=0.12$.}
		\label{Fig:R0_Rosenbrock}
	\end{figure}

An investigation of this influence on accuracy is performed through comparison of the $\ell_2$ error of the metamodel as a function of both the total
number of samples and the value of $R_0$. Figure \ref{Fig:R0_Rosenbrock} demonstrates the influence of this parameter when interpolating the
Rosenbrock function above. The comparison was repeated for multiple test functions of varying linearity, all of which may be found in \ref{app:TestFunc}.

Table \ref{tab:R0} shows the $\ell_2$ error after adding $0$, $8$ and $80$ points to a $5 \times 5$ full factorial design through the DoE algorithm.
The functions used are provided in \ref{app:TestFunc}. Data for $R_0 = 0.5$ and $R_0 = 1.4$ are shown as a sample, though data on the range $R_0 =
[0.25,1.75]$ was collected.

In general, smaller values of $R_0$ work well when the underlying function has more localized extrema, but quickly result in a clumped samples which
perform poorly in general. A larger value of $R_0$ naturally emphasizes more of the space-filling criteria, and thus, performs more consistently than
small $R_0$ tests, but with fewer benefits over using a non-adaptive approach. This is why, for the functions in table \ref{tab:R0}, the $R_0=1.4$
designs generally, but not always perform better than the $R_0=0.5$ ones. Values in the range $R_0=[1,1.5]$ perform most consistently overall, which
is consistent with the results of \cite{Mackman2010}. A value of $1.25$ is used throughout the remainder of this work.

	\begin{table}[h!]
	\centering
		\begin{tabular}{c c c c c c}
		\hline \hline
		\textbf{Function} & & \multicolumn{2}{ c }{$R_0 = 0.5$} & \multicolumn{2}{ c }{$R_0 = 1.4$} \\
		 & 0 Points & 8 Points & 80 Points & 8 Points & 80 Points \\
		\hline
		T1 Gaussian & 4.66 & 1.56 & 0.30 & 0.76 & 0.01\\ 
		T2 Rosenbrock & 7.84 & 7.27 & 5.13 & 7.10 & 4.04 \\ 
		T3 & 9.32 & 8.80 & 1.54 & 6.91 & 0.63 \\ 
		T4 & 8.31 & 7.83 & 6.50 & 7.60 & 3.20 \\ 
		T5 & 34.93 & 29.77 & 15.42 & 32.79 & 11.71 \\ 
		T6 & 16.75 & 9.40 & 3.82 & 10.57 & 2.21 \\ 
		T7 Branin & 1.44 & 1.38 & 0.60 & 1.47 & 0.57 \\ 
		T8 Himmelblau & 8.78 & 7.615 & 2.70 & 7.48 & 1.95 \\ 
		T9 Rosenbrock & 10.57 & 8.62 & 4.59 & 9.46 & 5.52 \\ \hline \hline 
		\end{tabular}
		\caption{$\ell_2$ error ($\time 10^{2}$) at $8$ and $80$ points for several test functions detailed in \ref{app:TestFunc}. Designs produced with a spacing parameter of
		$R_0=0.5$ and $R_0=1.4$ are compared. HOLMES approximants computed using $p=3$, order 3, and $\gamma=0.12$.}
		\label{tab:R0}
	\end{table}

	\subsection{Influence of Error}
	\label{subsec:InfError}

Comparing the proposed method to LHS or any Monte Carlo-style simulation in the realm of robustness against noise is not especially helpful for
assessing an adaptive DoE approach. Randomized simulations with an equal chance of choosing any point within the domain will naturally be the least
statistically biased, and therefore the most robust against errors in the data. A better comparison for the DoE method in general is to a similar
adaptive sampling method designed for unreplicated experiments.

An adaptive DoE method with the objective function given by $S(x) = Q_{L} Q_{S}$ was previously found to yield strong results on a number of test
functions \cite{Mackman2010}. This objective function uses an identical spacing parameter, $Q_S$. The linearity criterion is also similar, but created
using RBF interpolation with C2 Wendland functions rather than HOLMES functions. There is no LOO error criteria. The linearity criterion as originally
provided was not normalized onto $[\epsilon,1]$, but has been here to ensure more consistent performance than the original algorithm, which was found
to perform erratically on most functions.

\begin{figure}[h]
	  \centering
	  \includegraphics[width=0.66\textwidth]{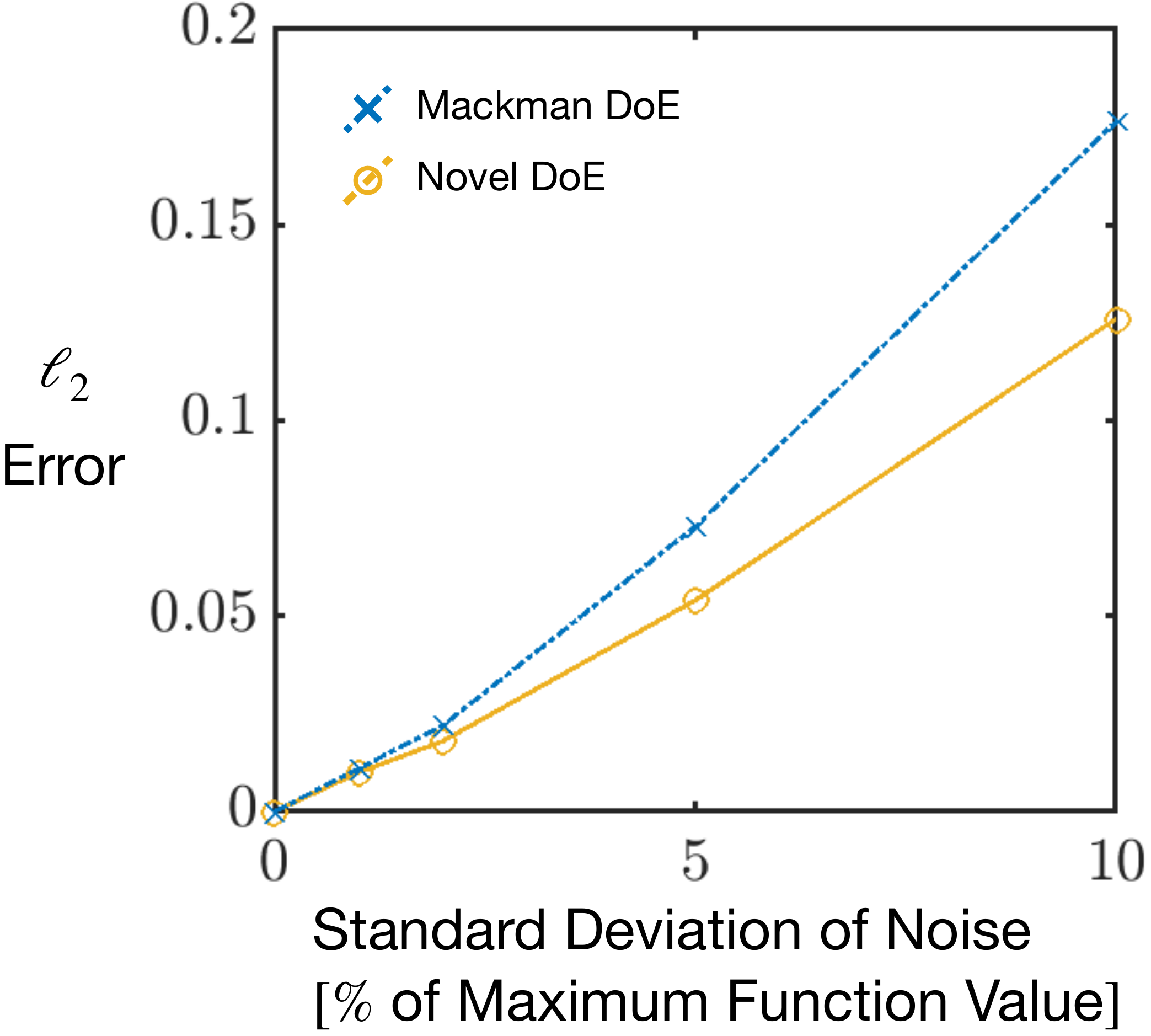}
	  \caption{Comparison of Novel DoE and Mackman DoE with regards to robustness in data noise and $36+40\times 4$ data points. $\ell_2$ error in
		plane function interpolation is calculated with $100\times 100$ points with different standard deviations of data point noise,
		measured as a percentage of the maximum function magnitude. HOLMES approximants computed using $p=3$, order 3, and $\gamma = 0.12$.
		Both algorithms use a nodal spacing parameter $R_0=1.25$.}
	  \label{fig:DoEError}
	\end{figure}

Let us again consider a plane function, but apply proportional Gaussian error to the collected data at each iteration. Both DoE algorithms are then
run with $36 + 40\times4$ data points. Since \autoref{subsec:HOLMESNoise} demonstrated the superiority of HOLMES in direct interpolation with noise,
it is not fair to compare the quality of data point placements using different interpolation methods for final error calculations. Instead, after all
points are collected, the solution is interpolated via a separate interpolation scheme -- LME interpolants -- and the $\ell_2$ error is calculated on
a $100 \times 100$ grid over the domain $[-1,1]^2$. This ensures the quality of the DoE point placements is compared, and not just the final
interpolation method. The results are seen in \autoref{fig:DoEError}. The difference between the two methods is insignificant for small amounts of
error, but becomes very noticeable once a moderate amount of error is achieved.
higherhigher

The improvement in accuracy does come at a computational cost. In the tests performed here, HOLMES functions generally evaluate slower than an
equivalent RBF interpolation; however, RBF interpolation requires inverting a matrix whose size depends on the number of data points, while HOLMES
interpolation scales instead with the number of points the metamodel is evaluated at. There could exist a situation in which this difference in
scaling proves favourable for the HOLMES DoE. Nonetheless, in most cases custom shape functions like HOLMES come at a computational cost. This cost
can be minimized or ignored if data collection is much slower than the DoE algorithm, or if computation of HOLMES shape functions can be smoothly
integrated into a computer experiment itself. For example, by choosing HOLMES as the shape function in a scheme for solving PDEs -- a topic which will
be briefly considered in the next section.

	\subsection{Adaptive DoE for Time Dependent Functions}
	\label{subsec:ADoEforTimeFuncs}

Time dependent functions are not generally considered as distinct from other functions in DoE methodology. Time may be considered a separate dimension
in the design space, and points placed according to that. This approach makes sense in a non-adaptive method, since all points can be established
prior to the start of testing. For an adaptive method this cannot work, since the experimentalist cannot collect data from a time that has already
passed.

The top-level approach taken here is very basic: known data points from previous times are projected to some point in the future, and the metamodel is
constructed with these points. This projected metamodel is then used in the DoE approach as described above. New points are proposed, and these are
used at the next timestep. In addition to a metamodel without any sort of projection, the approximation methods used are forward Euler (FE), where
$u(x,t_{n+1}) \approx u(x,t_n) + \sfrac{\partial u}{\partial t}|_{t_n}\Delta t$, backward Euler (BE), where $u(x,t_{n+1}) \approx u(x,t_n) +
\sfrac{\partial u}{\partial t}|_{t_{n+1}}\Delta t$, and explicit midpoint method (ME), where $u(x,t_{n+1}) \approx u(x,t_n) + \sfrac{\partial
u}{\partial t}|_{t_{n+1/2}}\Delta t$ and $u(x,t_{n+1/2}) \approx u(x,t_n) + \sfrac{\partial u}{\partial t}|_{t_{n}}\sfrac{\Delta t}{2}$. All
derivatives with respect to $t$ are estimated via a second order, three-point, backwards difference approximation when applicable, but since the
recently added nodes did not necessarily exist in the past, their values are estimated based on previous metamodels, which are stored for this
purpose.

Another important detail of the approach is the presence of a set of nodes arranged in a coarse full factorial grid which remains throughout all
timesteps. In the simulations considered, a constant $5 \times 5$ grid of nodes will exist in every timestep, with the remaining $24$ nodes added
according to the DoE algorithm at each timestep. The results are compared to a full factorial design of $7 \times 7$ nodes.

The axisymmetric $2d$ wave equation,
	\begin{gather} \label{eqn:Wave2d}
	  u_{tt} = c^2 \left( \frac{1}{r}\left(ru_{r}\right)_r + \frac{1}{r^2}u_{\theta \theta} \right),
	\end{gather}
is chosen for investigating this approach. The solution will produce a smoothly time-evolving function, and is relevant to both physical and computer
experiments. Specifically, the classical problem of vibration on an axisymmetric circular drum of radius $R=\sqrt{2}$ and constant $c=1$ is chosen,
resulting in boundary conditions of $u(R,t)=0$ and regularity in the solution at $u(0,t)$. A Gaussian initial condition is selected as
	\begin{gather} \label{eqn:WaveICs}
	  u(r,0) = U\left( e^{-\frac{\xi}{R^2}r^2} - e^{-\xi} \right).
	\end{gather}

The solution can be found via separation of variables as \eqref{eqn:WaveSolution}.
	\begin{gather} \label{eqn:WaveSolution}
	  u(r,t) = \sum_{n=1}^{\infty} A_n \cos \left(c \lambda_{0n}t \right) J_0\left(\lambda_{0n}r\right),
	\end{gather}
where $\lambda_{0n} = z_{0n}/R$, and $z_{0n}$ is the $n^{th}$ zero of the Bessel function of the first kind, $J_0(z)$. The coefficients, $A_n$, are
determined via orthogonality. The summation requires 33 terms to match the initial condition in \eqref{eqn:WaveICs} to computer precision.

\begin{figure}[h]
	  \centering
	  \includegraphics[width=\textwidth]{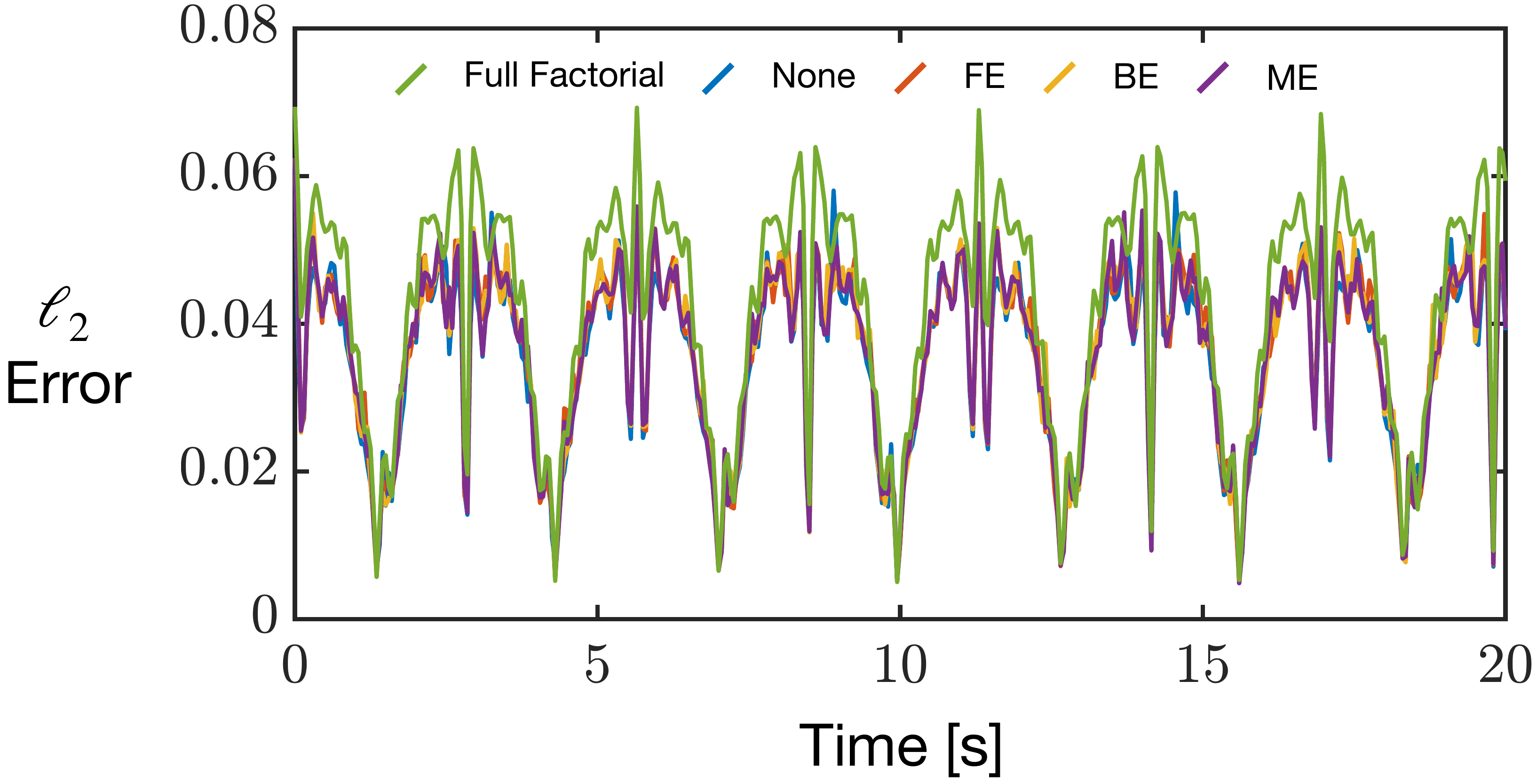}
	  \caption{Discrete $\ell_2$ error in metamodel as a function of time over $[-1,1]^2$ with $50 \times 50$ sample points. HOLMES approximants
	    computed using $p=3$, order 3, and $\gamma = 0.12$. DoE methods used are as follows: (a). Constant full factorial design (FF), and adaptive
	    DoE algorithm with nodes projected into the future using (b). No projection (None), (c). Forward Euler (FE), (d). Backward Euler (BE), and
	    (e). Explicit Midpoint method (ME).}
	  \label{fig:WaveL2vTime}
	\end{figure}

For convenience, the resulting metamodel accuracy is evaluated on an evenly spaced $100 \times 100$ grid of points covering $[-1,1]^2$. The nature of
the problem investigated results in error norms which oscillate in time, as in \ref{fig:WaveL2vTime}, so direct comparison between methods becomes
difficult, other than to note the DoE methods seem to provide modest, though visible increases in accuracy over the full factorial design regardless
of the method used to project node values through time.

\begin{figure}[h]
	  \centering
	  \includegraphics[width=0.66\textwidth]{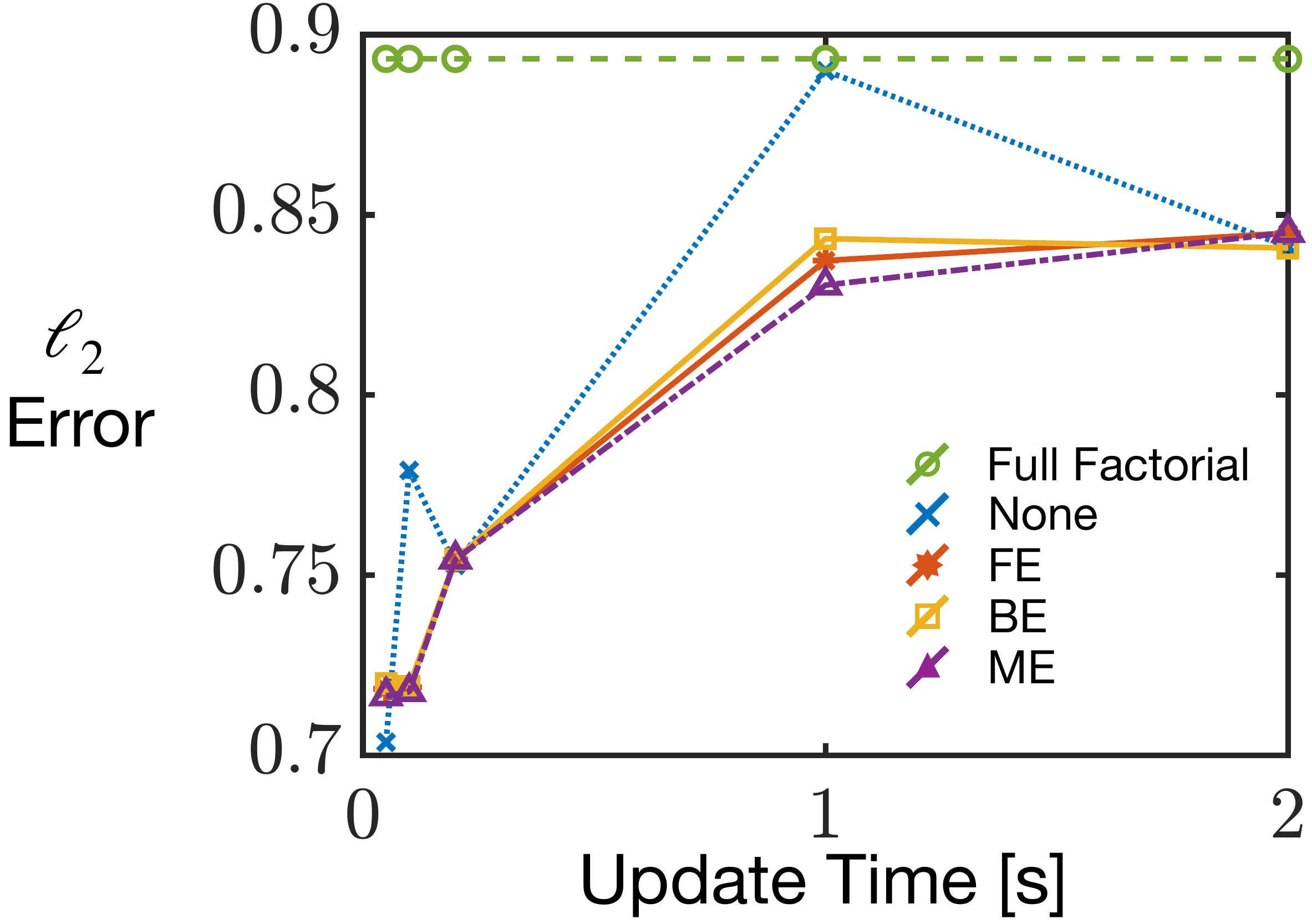}
	  \caption{Discrete $\ell_2$ error in metamodel as measured over $[-1,1]^2$ with $50 \times 50$ sample points integrated over the time
	    interval $[0,20]$s vs DoE update interval. HOLMES approximants computed using $p=3$, order 3, and $\gamma = 0.12$. DoE methods used are
	    as follows: (a). Constant full factorial design (FF), and adaptive DoE algorithm with nodes projected into the future using (b). No
	    projection (None), (c). Forward Euler (FE), (d). Backward Euler (BE), and (e). Explicit Midpoint method (ME).}
	  \label{fig:WaveL2vTimestep}
	\end{figure}

A quantitative comparison is easier if the error is treated as an integrable function in time, and integration is performed numerically. In this case,
trapezoidal rule is used over a simulation time of $20$s. The results in \ref{fig:WaveL2vTimestep} conform to expectations. Namely, some reasonable
increase in accuracy is generally achieved by the DoE methods compared to the full factorial approach. The increase is around $10-15\%$ when DoE
method is allowed to update the node positions frequently, but decreases as node placements become less frequent. For long update times, the
difference between DoE and full factorial performance becomes insignificant.

The projection of DoE nodes in the metamodel does have some effect in mitigating the influence of less frequent node placments. Un-projected nodes
result in a DoE method which is most sensitive in this regard, while all the projected methods appear to be more robust. Performance of all methods
becomes worse with decreased DoE placement frequency.

The approach described in this section is simple, and has some obvious limitations. It only makes sense to project data points a small time into the
future, and clearly the efficacy of the resulting DoE is dependent on the accuracy of the approximation. The allowable timestep will depend on the
experiment at hand, and the experimentalist will likely require some \textit{a priori} knowledge of the timescale of their problem. In the above
example, knowledge of the analytical solution was used to approximately set a suitable timestep, and the results were satisfactory without additional
effort.

The density of the initial full factorial grid could also be examined in detail. Would a larger proportion of points added through the
adaptive DoE method be beneficial, or detrimental? No attempt is made to investigate this beyond noting that, for the arbitrary combination of full
factorial and DoE points chosen, there was some improvement without any attempt at tuning, which could hint at some robustness in the approach.

Additionally, in a physical experiment the experimentalist must have time to both run the DoE algorithm, and adjust any measurement devices to capture
the recommended data points, within the time constraints posed by the forward projection of data points. It would be unsurprising if such conditions
could not be met in the vast majority of physical experiments; however, computer experiments would be much more likely to benefit from such an
approach. A time-dependent computer experiment could be periodically paused for the insertion of additional points, and resumed as needed -- just as
is done in the above example, but where the additional points are used in the experimental calculations. This exploration is out of scope of the
current work, but it is possible that particle-based continuum mechanics simulations, such as smoothed particle hydrodynamics, could benefit from such
an approach. Errors will accumulate through time in such simulations, so the cumulative effect of even modest increases in accuracy can be significant
by the end of a simulation.


\section{Conclusion}
\label{sec:Conclusion}

We proposed a novel adaptive DoE method that balances curvature, space filling, and metamodel error considerations to automatically select nodes with
an optimal location in the domain and approximate non-linear functions within the design space. HOLMES with a variable kernel size is introduced to
satisfy the adaptive nature of the proposed method. Adaptive HOLMES metamodelling provides accurate function and derivative approximations on the
interior of the domain, making it more suitable for handling data with moderate Gaussian white noise than schemes using radial basis functions. The
key parameter of the method, $R_0$, was investigated and can generally be set within a limited range. Additionally, we proposed alternative ways to
estimate the local nodal spacing in unstructured data using the boundary-corrected KDE method.

Evaluation of the new method indicated better performance than simple, established, nonadaptive methods such as LH and FF on a set of test functions.
Then method outperforms a similar adaptive DoE method when noise is introduced into the simulation. Additionally, the approach becomes more applicable
to time-dependent experiments when the metamodels were first projected forward in time.

Future work could adapt the proposed approach to the solution of partial differential equations. Specifically, this approach's unstructured,
multidimensional nature could lend itself to particle-based continuum mechanics models, such as smoothed particle hydrodynamics, which are still
developing strategies for dynamically and methodically increasing resolution during the simulation.

\section{Acknowledgment}
We acknowledge the support from the Natural Sciences and Engineering Research Council of Canada (NSERC) through the Discovery Grant under Award Application number 2016-06114 and the Alliance grant number AWD-014405. L.C. acknowledges the financial support from NSERC CGS M fellowship. 
This research partially was supported through computational resources and services provided by Advanced Research Computing at the University of British Columbia and Compute Canada.

\appendix
\section{Test Functions}
\label{app:TestFunc}

The analytical test functions used were as follows. All were normalized onto $[-1,1] \times [-1,1]$. To compare $\ell_2$ norm values between
functions, all functions are also scaled by a constant $A$ such that $\max_{x\in [-1,1]^2} |f(x)| = 1$.

	\begin{itemize}
		\item[T0] Plane: $f(x,y) = x + y$, \\$x,y \in [-1,1]$.
		\item[T1] Gaussian Hill: $f(x,y) = e^{-3(x^2 + y^2)}$, \\$x,y \in [-1,1]$.
		\item[T2] Log-Scaled Rosenbrock: $f(x,y) = \log \left( 1 + (100(y-x^2)^2 + (1-x)^2 \right)$, \\$x,y \in [-1,1]$.
		\item[T3] $f(x,y) = 2 + 0.01(y-x^2)^2 + (1-x)^2 + 2(2-y)^2 + 7\sin(0.5x)\sin(0.7xy)$, \\$x,y \in [0,5]$.
		\item[T4] $f(x,y) = \cos(6(x-\frac{1}{2})) + 3.1(|x-0.7|) + 2(x-\frac{1}{2}) + \sin\left(\frac{1}{|x-\frac{1}{2}|+0.31}\right) + \frac{y}{2}$, \\$x,y \in [0,1]$.
		\item[T5] $f(x,y) = \cos \left((x^2 + y^2)^{1/2} \right)$, \\$x,y \in [-5,5]$.
		\item[T6] $f(x,y) = \sin(x)\sin(y)$, $x,y \in [-3,3]$.
		\item[T7] Branin's Function: $f(x,y) = \left(y - 5.1\left(\frac{x}{2\pi}\right)^2 + \frac{5x}{\pi} - 6\right)^2 + 10\left(1-\frac{\pi}{8}\right)\cos(x) + 10$, \\$x \in [0,15]$, $y \in [-5,10]$
		\item[T8] Himmelblau's Function: $f(x,y) = (x^2 + y - 11)^2 + (x + y^2 - 7)^2$, \\$x,y \in [-4,4]$.
		\item[T9] Rastrigin's Function: $f(x,y) = 20 + x^2 + y^2 - 10(\cos(2\pi x) + \cos(2\pi y))$, \\$x,y \in [-1,1]$.
	\end{itemize}

\bibliographystyle{unsrt}
\bibliography{DoE_with_HOLMES.bib}

\end{document}